\documentclass[12pt]{oxarticle}
\pdfoutput=1
\usepackage{amsmath, amssymb}
\usepackage{graphicx}
\usepackage{kbordermatrix}

\let\SS=\S 
\renewcommand{\a}{\alpha}
\renewcommand{\b}{\beta}

\newcommand{\z}{\zeta}

\renewcommand{\k}{\kappa}
\renewcommand{\l}{\lambda}
\newcommand{\m}{\mu}
\newcommand{\n}{\nu}
\newcommand{\x}{\xi}

\newcommand{\p}{\pi}
\renewcommand{\r}{\rho}
\renewcommand{\S}{\Sigma}
\renewcommand{\t}{\tau}

\newcommand{\ph}{\phi}\newcommand{\vph}{\varphi}
\newcommand{\ch}{\chi}

\renewcommand{\o}{\omega}

\newcommand{\cB}{\mathcal{B}}
\newcommand{\cC}{\mathcal{C}}

\newcommand{\cF}{\mathcal{F}}

\newcommand{\cN}{\mathcal{N}}
\newcommand{\cO}{\mathcal{O}}

\newcommand{\cT}{\mathcal{T}}

\newcommand{\cV}{\mathcal{V}}
\newcommand{\cW}{\mathcal{W}}

\newcommand{\IC}{\mathbb{C}}

\newcommand{\IH}{\mathbb{H}}

\newcommand{\IP}{\mathbb{P}}
\newcommand{\IQ}{\mathbb{Q}}
\newcommand{\IR}{\mathbb{R}}

\newcommand{\IZ}{\mathbb{Z}}

\font\twentyfourrm=cmr12 at 24pt
\font\eightrm=cmr8 at 8pt
\font\csc=cmcsc10

\newcommand{\beq}{\begin{equation}}
\newcommand{\eeq}{\end{equation}}
\newcommand{\bea}{\begin{eqnarray}}
\newcommand{\eea}{\end{eqnarray}}
\newcommand{\bean}{\begin{eqnarray*}}
\newcommand{\eean}{\end{eqnarray*}}
\newcommand{\eref}[1]{(\ref{#1})}

\newcommand{\comment}[1]{}
\newcommand{\nn}{\nonumber}
\newcommand{\pd}[2]{\frac{\partial #1}{\partial #2}}
\newcommand{\diff}[1]{\frac{\partial}{\partial #1}}
\newcommand{\vertequals}{\def\shortvr{\vrule height 5pt depth3pt width0.3pt}\shortvr\hskip2pt\shortvr}

\newcommand{\defineas}{=}

\newcommand{\cy}{Calabi--Yau}
\newcommand{\cym}{Calabi--Yau manifold}
\newcommand{\cys}{Calabi--Yau manifolds}
\newcommand{\hodgenos}{(h^{11},\,h^{21})}
\newcommand{\quotient}[1]{_{\hskip-2pt\lower1pt\hbox{$/$}\lower2pt\hbox{\hskip-1pt$#1$}}}

\newcommand{\Vt}{\ensuremath{\widetilde{\cV}}}
\newcommand{\ZZ}{{\ensuremath{\IZ_3\times\IZ_3}}}
\newcommand{\zz}{{\ensuremath{\IZ_3{\times}\IZ_3}}}
\newcommand{\fref}[1]{Figure~\ref{#1}}
\newcommand{\tref}[1]{Table~\ref{#1}}
\newcommand{\squig}{\leftrightsquigarrow}
\newcommand{\End}{\mbox{End}}
\newcommand{\cicystop}{~\lower8pt\hbox{.}}

\newcommand{\Bigcheck}{\lower3.8pt\hbox{\smash{\hbox{{\twentyfourrm \v{}}}}}}
\newcommand{\Mcheck}{\kern4.4pt\hbox{\Bigcheck\kern-13.5pt{$M$}}}
\newcommand{\cicy}[2]{\begin{matrix} #1\end{matrix}\!\left[\begin{matrix}#2 \end{matrix}\right]}
\newcommand{\Mone}{\cicy{\IP^{3}\\ \IP^{3}\\}{1&3&0\\ 1&0&3\\}_{\hskip-3pt -18}}
\newcommand{\Mtwo}{\cicy{\IP^{2}\\ \IP^{3}\\}{3&0\\ 1&3\\}_{\hskip-3pt -54}}
\newcommand{\Mthree}{\cicy{\IP^{2}\\ \IP^{2}\\}{3\\ 3\\}_{\hskip-3pt -162}}

\proofmodefalse
\begin{document}
\pagestyle{empty}
\begin{center}
\null\vskip0.3in
{\Huge Triadophilia$^{\hbox{\footnotesize *}}\hskip2pt$:\\[1ex]
A Special Corner in the Landscape\\[0.5in]}
{\csc Philip Candelas$^1$,~ Xenia de la Ossa$^1$,~ Yang-Hui He$^{1,2}$\\[0.5ex]
and\\[0.5ex]
Bal\'{a}zs Szendr\H{o}i$^1$\\[0.5in]}
{\it $^1$Mathematical Institute\hphantom{$^1$}\\
Oxford University\\
24-29 St.\ Giles'\\
Oxford OX1 3LB, England\\[4ex]
$^2$Merton College\hphantom{$^2$}\\
Oxford OX1 4JD, England}
\end{center}
\vfill
\centerline{\bf Abstract}
\parbox{6.5in}{\setlength{\baselineskip}{14pt}
It is well known that there are a great many apparently consistent vacua of string theory. We draw attention to the fact that there appear to be very few Calabi--Yau manifolds with the Hodge numbers $h^{11}$ and           
$h^{21}$ both small.  Of these, the case $(h^{11}, h^{21})=(3,3)$ corresponds to a manifold on which a three generation heterotic model has recently been constructed. We point out also that there is a very close relation between this manifold and several familiar manifolds including the `three-generation' manifolds with 
$\chi=-6$ that were found by Tian and Yau, and by Schimmrigk, during early investigations. It is an intriguing possibility that we may live in a naturally defined corner of the landscape. The location of these three generation models with respect to a corner of the landscape is so striking that we are led to consider the possibility of transitions between heterotic vacua. The possibility of these transitions, that we here refer to as transgressions, is an old idea that goes back to Witten. Here we apply this idea to connect three generation vacua on different \cy~manifolds.}
\renewcommand{\thefootnote}{\fnsymbol{footnote}}
\footnotetext[1]{~{\it Triadophilia\/} from G.\ a love of three-ness, a nostalgia for a world of three generations. Less precise but also less cumbersome than {\it tritogeneia-philia}.}
\renewcommand{\thefootnote}{\arabic{footnote}}
\newpage
\tableofcontents
\newpage
\setcounter{page}{1}
\pagestyle{plain}
\section{Introduction and Summary}
\subsection{Survey of constructions of \cys.}
Until the interest in Calabi--Yau manifolds that derived from string theory, very few of these were known explicitly; the manifolds had, at that time, only recently been shown to exist. Owing to the interest from string theory, increasingly large classes of Calabi--Yau manifolds were constructed. Since, however, it seems to be impossible  to construct classes of manifolds with desired properties, one must perforce construct large classes of manifolds that are then searched for cases that are phenomenologically interesting. 
Tian and Yau~\cite{TianYau} were nevertheless able, at a very early stage, to find two manifolds with Euler number $\chi=-6$ leading to a model with three generations of particles.
Both of these have Hodge numbers $(h^{11},h^{21})=(6,9)$. Motivated by these examples            Schimmrigk~\cite{Schimmrigk} found a third manifold with $\chi=-6$, and with the same Hodge numbers. This same manifold was rediscovered, shortly afterwards, by Gepner~\cite{Gepner} in the process of constructing rational conformal field theories. We can denote the three families of manifolds, families because they have parameters, in the following way:
\beq
M''=N''/A~,~~~M'=\widehat{N'/A{\times }B} ~~\mbox{and}~~
M=\widehat{N/A{\times} B{\times} C}~,
\label{quotients}
\eeq
where
\beq
N''~=~\Mone~,~~
N'~=~\Mtwo~,~~
N~=~\Mthree~.
\label{threedefs}
\eeq
The notation denotes that $N''$, for example, is realised in the product $\IP^3{\times}\IP^3$ by three equations whose degrees, in the variables of the two projective spaces, are given by the columns of the matrix.
The subscripts appended to the matrix are the Euler numbers of the manifolds and the notation in~\eref{quotients}
indicates that the manifolds are quotiented by certain groups $A$, $B$ and $C$. Each of these groups is abstractly a  $\IZ_3$. The group $A$ acts freely but $B$ and $C$ each leave fixed a certain curve, in fact a torus, within the manifold and the hats indicate that these fixed tori are resolved. It was suspected, on the basis of the identity of the Hodge numbers together with the fact that they fall into a sequence, that the manifolds in fact belong to the same irreducible family, and this was shown to be the case in \cite{GreeneKirklin}.

Despite the ease with which these early examples had been found, further examples of manifolds with $\chi=\pm 6$ proved much more elusive. The class of all \cys\ that can be realised as a complete intersection of polynomials in a product of projective spaces, hence known as CICY's, generalising the construction  of the manifolds of~\eref{threedefs}, was constructed in~\cite{CICYsI, CICYsII, OxfordCICYs}. This class, consisting of almost 8,000 manifolds, was searched for manifolds with $\chi=-6$, and for manifolds whose quotients by a freely acting group could have $\chi=-6$. None was found beyond the three above which inspired the construction. 

\begin{figure}[!t]
  \begin{center}
\includegraphics[width=6in]{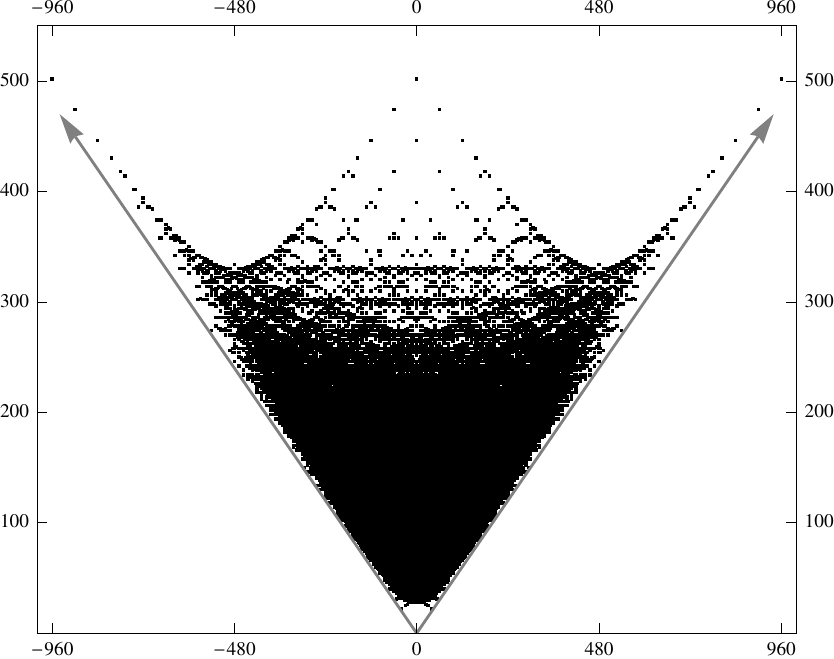}
\vskip10pt
\parbox{4.9in}{\caption{\label{bigplot}
\small A plot of the Hodge numbers of the Kreuzer--Skarke list. 
\hbox{$\ch=2(h^{11}-h^{21})$} is plotted horizontally and $h^{11}+h^{21}$ is plotted vertically. The oblique axes bound the region $h^{11}\geq 0,\,h^{21}\geq 0$.}}
\end{center}
\end{figure}

The number of examples of \cys\ was increased by the construction of 
manifolds~\cite{CandelasLynkerSchimmrigk, KlemmSchimmrigk, KreuzerSkarkeWeighted} given by polynomials in weighted $\IP^4$ and increased again very greatly by the construction of manifolds as hypersurfaces in toric varieties following the methods introduced by Batyrev~\cite{Batyrev}. In a tour de force of computer calculation \cite{Kreuzer:2002uu, KreuzerSkarkeReflexive} Kreuzer and Skarke compiled a list of all four-dimensional reflexive polyhedra, each of which corresponds to a family of anticanonical hypersurface \cys\ in the corresponding toric variety. The list runs to almost 500,000,000 polyhedra and gives rise to some 30,000 distinct pairs of Hodge numbers\footnote{It is not known how many of these manifolds are distinct. Manifolds with distinct Hodge numbers are certainly distinct, however the converse is not true in general, so the number of distinct manifolds is somewhere between 30,000 and 500,000,000. For the CICY's there are 264 pairs of Hodge numbers and roughly 8,000 manifolds. For this case it is known~\cite{HeCandelas} that at least 2590 of the manifolds are distinct as classical manifolds.} which we plot in \fref{bigplot}. 

\begin{figure}[!t]
  \begin{center}
\includegraphics[width=6in]{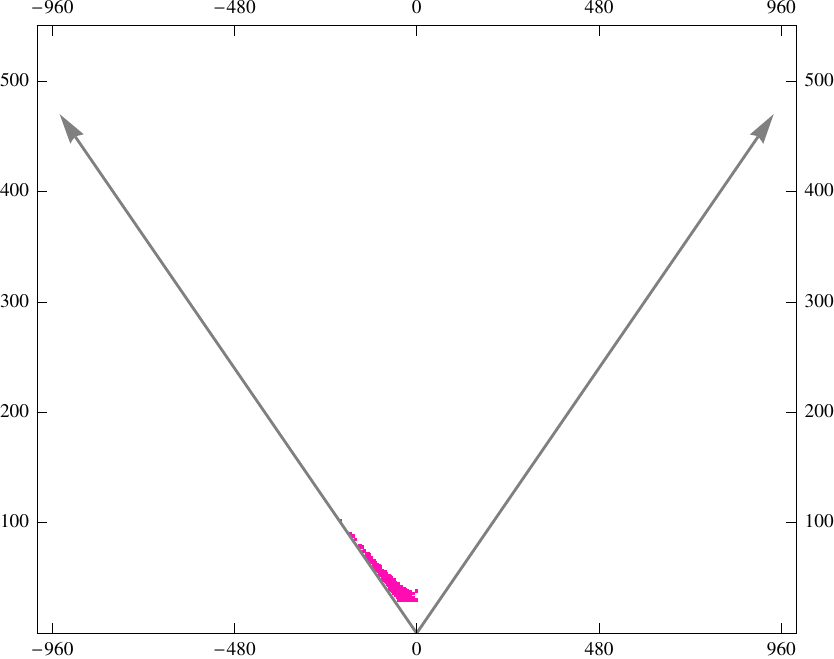}   
  \end{center}
\caption{\label{cicyplot}\small A plot of the 264 distinct pairs of Hodge Numbers for the CICY's.}
\end{figure}

For comparison we include a plot in \fref{cicyplot} of the Hodge numbers of the 263 distinct pairs of Hodge numbers for the CICY's plotted to the same scale. This is something of a cautionary tale showing what can happen when a seemingly large class of manifolds turns out to be rather special.

The Kreuzer--Skarke list, vast though it is, does not exhaust all possibilities. An obvious extension is to include the possibility of higher codimension corresponding to the case of more than one polynomial in a toric variety of higher dimension; these one might term toric CICY's. A correspondence with lattice polyhedra that generalizes the construction of Batyrev for the case of a single polynomial has been given by Batyrev and Borisov~\cite{BatyrevBorisov}. Two simple examples of such manifolds will appear later and it is worth writing one of them here to explain the notation and to give an idea of the immense number of possible members of this class. Consider the manifold that is denoted by
$$
\IP\left(\begin{matrix}1\,1\,1\,1\,1\,1\,0\cr 1\,1\,0\,0\,0\,0\,1\cr \end{matrix}\right)\hskip-3pt
     \left[\begin{matrix} 3~3\cr 3~0\cr \end{matrix}\right]\cicystop
$$
The first matrix is the weight matrix and the second one is the degree matrix.
Each column of the weight matrix corresponds to a coordinate: so in this case we have coordinates 
$(z_1,\,z_2,\ldots,\,z_7)$ and the two rows of the first matrix indicate that there are two independent scalings with the columns of the matrix corresponding to the weights of each coordinate. Under a scaling the coordinates transform as
$$
(z_1,\, z_2,\, z_3,\, z_4,\, z_5,\, z_6,\, z_7)~\to~
(\l\m\,z_1,\, \l\m\,z_2,\, \l\,z_3,\, \l\,z_4,\, \l\,z_5,\, \l\,z_6,\, \m\,z_7)~,
$$
with $\l,\,\m$ nonzero complex numbers.
The second matrix indicates that there are two polynomials $p^1$ and $p^2$ and that under a scaling
$p^1\to \l^3\m^3\, p^1$ and $p^2\to\l^3\, p^2$. The fact that the manifold has vanishing first Chern class is ensured by the condition that the row sum of each row of the weight matrix is equal to the row sum of the corresponding row of the degree matrix. The dimension count, in this case, is that we have 7 coordinates that are identified under two scalings and subject to two polynomial constraints yielding a manifold of dimension
$7{-}2{-}2=3$. The question of when a configuration of this type gives rise to a nonsingular manifold is answered by the Batyrev--Borisov procedure. The Kreuzer--Skarke list corresponds to the special case that the degree matrix has only one column, the hypersurfaces in weighted $\IP^4$ correspond to the case that the weight matrix has only one row, and the 
CICY's to the very special case that all the entries of the weight matrix are either 1 or 0 and moreover that each column of the weight matrix contain precisely one 1. The number of possible configurations would seem to be immense and the scale of the enterprise of examining this class would seem to preclude any complete listing, though several hundred new pairs of Hodge numbers have been found by studying the interesting region along the edges of the plots where one of the Hodge numbers is 
small~\cite{KreuzerRieglerSahakyan, KlemmKreuzerRieglerScheidegger}. 

Batyrev and Kreuzer~\cite{BatyrevKreuzerConifolds} have also found many new pairs of Hodge numbers by examining reflexive polyhedra for hypersurfaces in toric varieties that admit conifold singularities, blowing down the $\IP^1$'s and smoothing the resulting manifolds.

Even this of course is not everything, since there are \cys\ that are not covered by these constructions; we are, moreover, also interested in heterotic vacua corresponding to vector bundles~$\cV$ on the \cy\ manifold for which $c_1(\cV)=0$. If \hbox{$\cV=\cT$}, the tangent bundle, then the number of generations is 
$\left|\frac{1}{2}c_3(\cT)\right|=\left|\frac{1}{2}\chi\right|$ so three generations corresponds to $\ch=\pm 6$ as is the case for $N=2$ compactifications. For general heterotic vacua, however, there is much greater freedom. The restrictions on $\cV$ are that it be stable, have $c_1(\cV)=0$ and have $c_2(\cV)$ satisfy a certain condition. The number of generations of particles is then $|\frac{1}{2}c_3(\cV)|$. There are presumably a great many of these heterotic vacua; while toric geometry has afforded us a considerable degree of control over \cys\, it is not yet known how to extend this degree of control to bundles over these manifolds. 

Special cases can, however, be studied and the group at the University of 
Pennsylvania~\cite{Braun:2004xv, Braun:2005nv, Donagi:2000zf, Bouchard:2005ag} has developed a small number of              
three-generation heterotic models based on quotients of a special\footnote{The split bicubic is special in so far as it has 
$(h^{11},h^{21})=(19,19)$ the value for $h^{11}$ being the largest for any CICY. All the CICY's have Euler numbers in the range $-200\leq\ch\leq 0$ and the split bicubic, together with a manifold with Hodge numbers
(15,15), are the only two CICY's which have $\chi=0$ and can possibly be self-mirror.} CICY, the split bicubic family
\beq
X^{19,19}~\defineas~~\cicy{\IP^{1}\\ \IP^{2}\\ \IP^{2}\\}{1&1\\3&0\\ 0&3\\}_{0}^{19,\,19} 
\label{splitbicubic}\eeq 
where we append superscripts to record the values of $\hodgenos$.
One special feature of the split bicubic is that it is a bi-elliptic fibration. To see this, consider the form of the equations~\cite{Braun:2007tp} for this space
\beq
\begin{split}
t_1\Big(\sum_{j=1}^3\x_j^3 -3a\,\x_1\x_2\x_3\Big)+ 3c t_2\,\x_1\x_2\x_3~&=~0 \\
3c t_1\,\eta_1\eta_2\eta_3  + t_2\Big(\sum_{j=1}^3\eta_j^3 - 3b\,\eta_1\eta_2\eta_3\Big)~&=~0~,
\nn\end{split}
\eeq
where we have chosen coordinates $t$ for the $\IP^1$ and $\x$ and $\eta$ for the two $\IP^2$'s. These are particularly symmetric polynomials of the given degrees of a form to which we will return later; the most general polynomials would contain 19 parameters. However this simple choice is sufficient to illustrate the following point. Consider the equations for fixed \hbox{$(t_1,t_2)\in\IP^1$;} each equation is then a cubic in a $\IP^2$, generically an elliptic curve (i.e., a two-torus). Thus the split bicubic is a fibration over $\IP^1$ with fiber $E_1(t){\times}E_2(t)$, where for generic $t$, both $E_i(t)$ are elliptic curves, which degenerate for certain special values of~$t$.

Holomorphic vector bundles on elliptic curves were classified by Atiyah~\cite{Atiyah} and the extension to spaces that are fibered by elliptic curves was considered by Donagi~\cite{elliptic} and by 
Friedman, Morgan and Witten~\cite{FMWI, FMWII}. Further work investigated the problem of constructing stable
$SU(n)$ bundles, for $n=3,4,5$, on the large class of Calabi--Yau threefolds that are elliptically fibered, that is for which there is a map to a $\IP^1$ for which the generic fiber is an elliptic 
curve~\cite{Andreas:1999ty, Donagi:2000zf}. An explicit construction of a heterotic model  whose low energy effective theory has the particle content of the Minimum Supersymmeteric Standard Model nevertheless proved elusive until such a model based on a stable $SU(4)$ vector bundle, corresponding to a gauge group $SO(10)$ in spacetime, was
presented in~\cite{Braun:2005nv}. The manifold of this model is a $\IZ_3{\times}\IZ_3$ quotient of the split bicubic. A breaking of the $SO(10)$ symmetry via the Hosotani mechanism, that takes advantage of the fundamental group $\IZ_3{\times}\IZ_3$, yields the particle spectrum of the MSSM, without exotics, and with no anti-generations. A version with $SU(5)$ vector bundle, hence also $SU(5)$ in spacetime, was also 
found~\cite{Bouchard:2005ag}. 

Returning to Figure~\ref{bigplot} and the Kreuzer--Skarke list, it is apparent that the central part of the plot is very dense with essentially every site occupied. The main point that we wish to make here is that the tip of the diagram where $(h^{11},h^{21})$ are both small is thinly populated and this remains true even if we include the CICY's, the Klemm--Kreuzer toric CICY's, the toric conifolds and other examples of which we are aware. One way of attempting to populate the tip is to seek \cys\ that are free quotients, with a nontrivial fundamental group. Such manifolds seem however to be genuinely rare and especially so for larger fundamental groups that would produce small Hodge numbers. This was apparent for the  CICY's from the first investigations~\cite{CICYsII, OxfordCICYs}. 

Recently Batyrev and Kreuzer
\cite{Batyrev:2005jc} have searched the Kreuzer--Skarke list for manifolds with a nontrivial fundamental group and find just 16 examples; moreover the fundamental groups that they find are: one occurrence of $\IZ_5$ and two occurrences of  $\IZ_3$, with the remaining 13 instances corresponding to $\IZ_2$'s. This is not everything: a quotient manifold will only appear in the Kreuzer--Skarke list if the quotient is realized torically. Thus the occurrence of the $\IZ_5$ corresponds to a quotient of the quintic threefold 
$$
\IP^4[5]\quotient{\IZ_5}~,~~~(h^{11},h^{21})=(1,\,21)$$
with the generator of the group corresponding to the action $x_j\to\z^j x_j$ on the coordinates of the embedding space, for $\z$ a nontrivial fifth root of unity. The further quotient
$$
\IP^4[5]\quotient{\IZ_5{\times}\IZ_5}~,~~~(h^{11},h^{21})=(1,\,5)$$
is not present in the list owing to the fact that the generator of the second symmetry group acts by cyclic permutation of the coordinates, not by multiplying the coordinates by roots of unity\footnote{It is possible to choose coordinates so that the first generator acts cyclically on the coordinates and the second acts by multiplication by fifth roots of unity. It is, however, not possible to arrange for both generators to act torically.}.
One of the two occurrences of $\IZ_3$ also involves a familiar manifold
$$
\cicy{\IP^{2}\\ \IP^{2}\\}{3\\ 3\\}\quotient{\IZ_3}~,~~~(h^{11},h^{21})=(2,\,29)$$
where the $\IZ_3$ can be chosen to be either the symmetry $A$ of~\eref{quotients} or a certain diagonal subgroup of $B{\times}C$. The other $\IZ_3$ quotient is 
$$
\IP(1\,1\,1\,3\,3)[9]\quotient{\IZ_3}~,~~~(h^{11},h^{21})=(2,\,38)~.
$$
The Batyrev--Kreuzer search does not find
$$
\cicy{\IP^{3}\\ \IP^{3}\\}{1&3&0\\ 1&0&3\\}\quotient{\IZ_3}~,~~~(h^{11},h^{21})=(6,\,9)$$
because this space is described by three polynomials while the Kreuzer--Skarke list corresponds to spaces that are defined by a single polynomial. 
 
\begin{figure}[!p]
\begin{center} 
\includegraphics[width=6.5in]{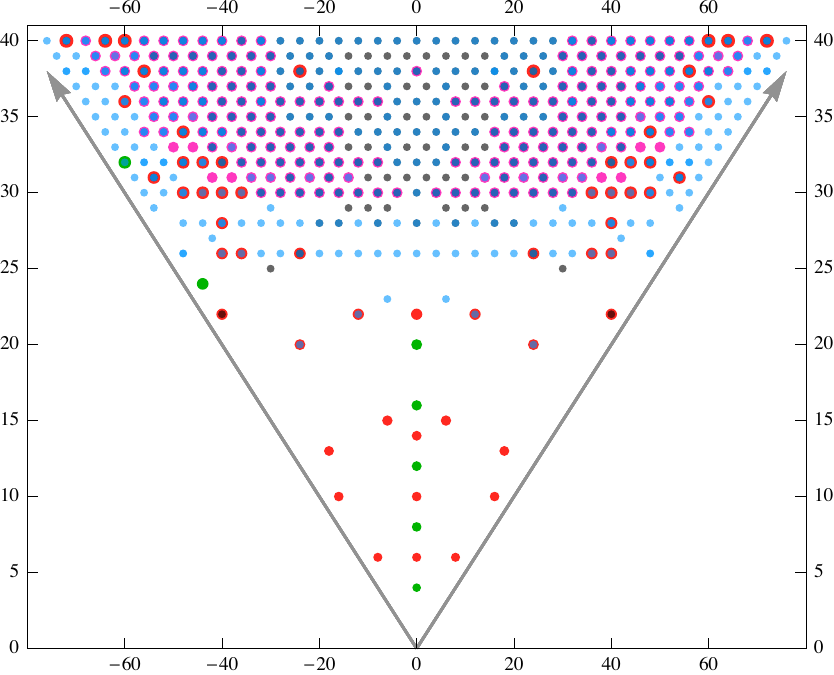}    
\vskip10pt
\parbox{5.5in}{\caption{\label{tipplot}
The underpopulated corner of the landscape. \hbox{$\ch=2(h^{11}{-}h^{21})$} is plotted horizontally,
$h^{11}{+}h^{21}$ is plotted vertically and the oblique axes bound the region 
$h^{11}\geq 0,\, h^{21}\geq 0$. In the electronic version of this figure the points are coloured according to provenance and have partial transparency in order to show overlays. The manifolds with 
$h^{11}{+}h^{21}\leq 22$ are identified in \tref{tiptab}.}}
\vskip20pt
\parbox{5in}{\small
\includegraphics[width=6pt]{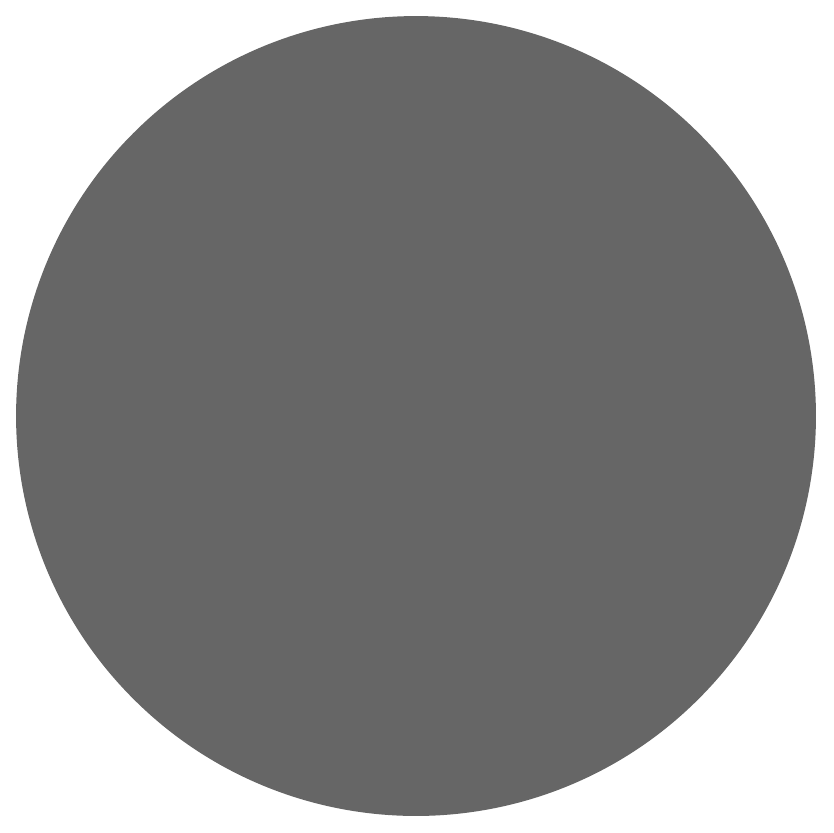}~~The Kreuzer--Skarke list.\\
\includegraphics[width=6pt]{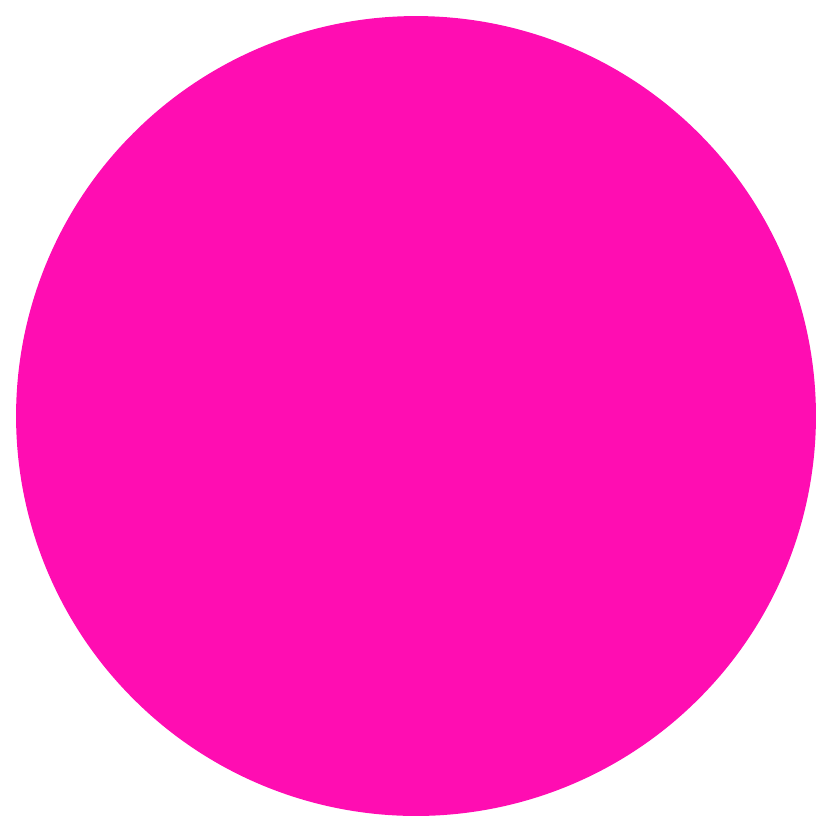}~~The CICY's and their mirrors.\\ 
\includegraphics[width=6pt]{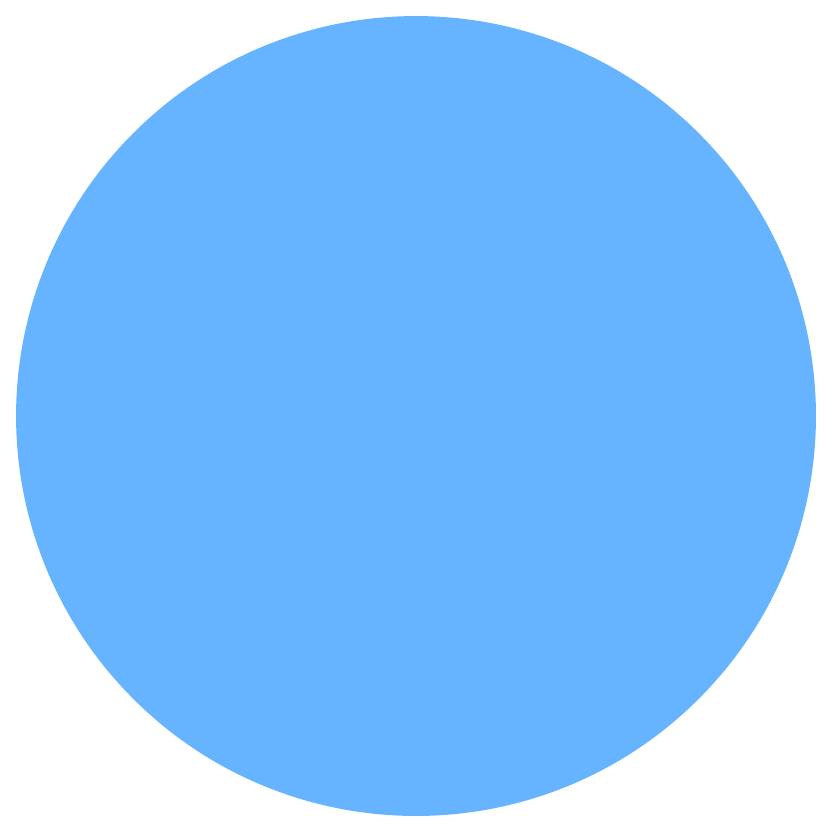}~~The toric CICY's together with the toric conifolds, and their mirrors. \\ 
\includegraphics[width=6pt]{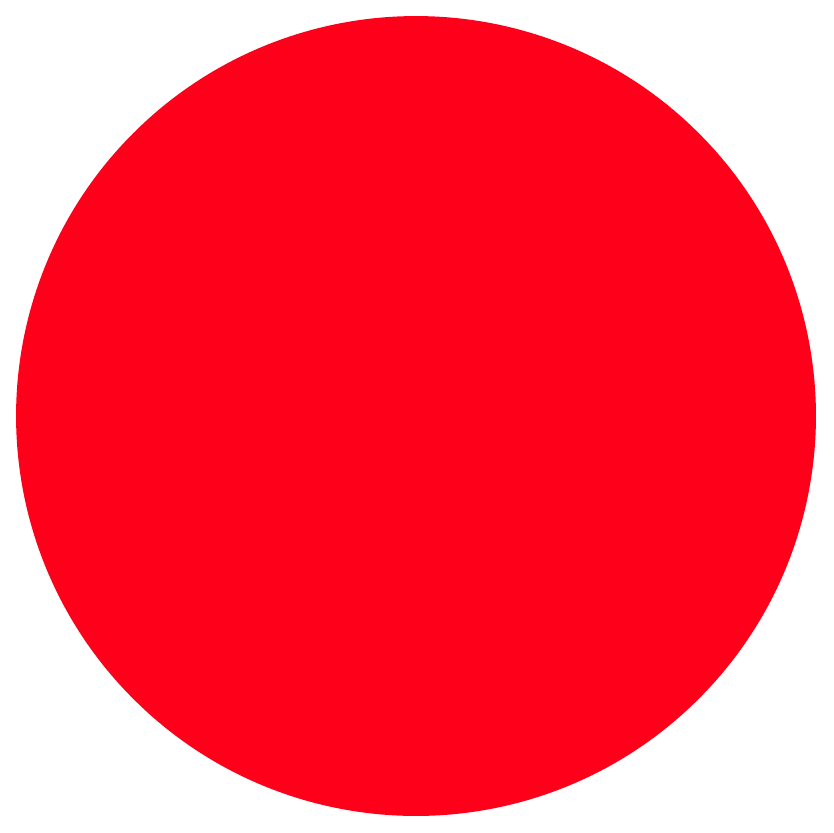}~~Quotients by freely acting groups and their mirrors.\\ 
\includegraphics[width=6pt]{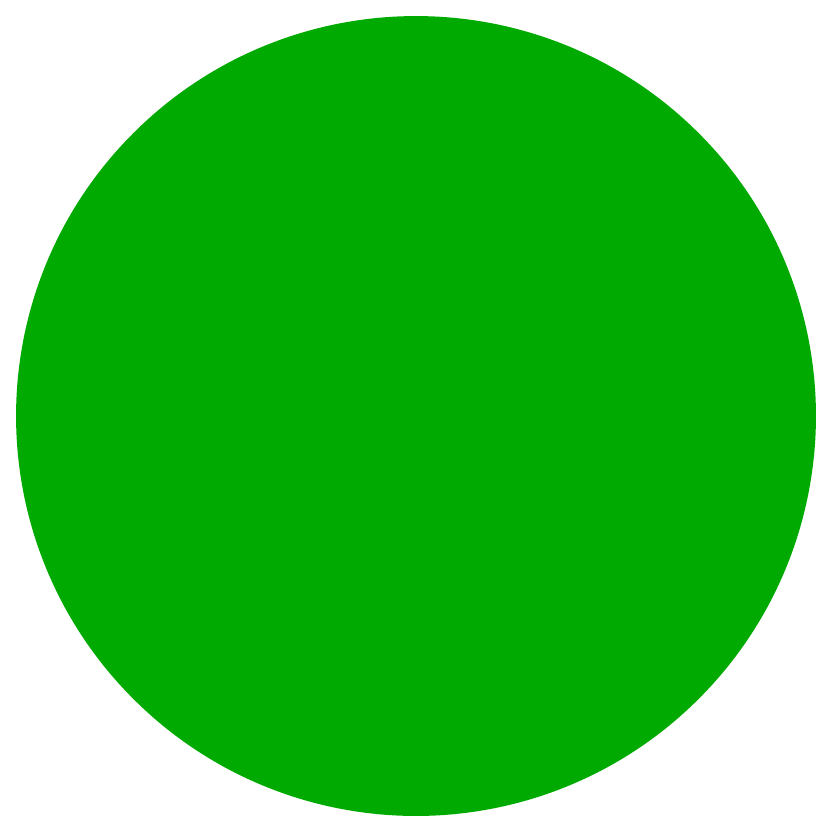}~~The Gross--Popescu and Tonoli manifolds.}
\end{center}
\end{figure} 
\begin{table}[p]
\def\str{\vrule height14pt width0pt depth8pt}
\def\Str{\vrule height20pt width0pt depth18pt}
\def\STR{\vrule height25pt width0pt depth24pt}
\def\SSTR{\vrule height35pt width0pt depth30pt}
\begin{center}
\begin{tabular}{| c | c | c | c |}
\hline \str $(\chi,\, y)$ & \hskip-4pt$(h^{11},h^{21})$\hskip-4pt\null & Manifold & Reference\\
\noalign{\hrule\vskip3pt\hrule}
\str (-40,22) & (1,21) & $\IP^4[5]/\IZ_5$ & --\\\hline
\STR (-12,22)&(8,14)&\footnotesize
$\IP\!\left(\begin{matrix} 4\,2\,2\,2\,1\,1\,0\,0\cr 0\,0\,0\,0\,1\,1\,2\,0\cr 1\,0\,0\,0\,0\,0\,0\,1\cr
		\end{matrix}\right)\hskip-5pt
	\left[\begin{matrix} 8~4 \cr 0~4 \cr 2~0 \end{matrix}\right]\quotient{(\IZ_2:1,1,0,0,1,0,1,0)}$
&\cite{KreuzerRiegler}\\\hline
\str (0,22) & (11,11) & $X^{19,19}/\IZ_2$ & \cite{BouchardDonagi}\\\hline
\Str (-24,20)& (4,16)&\footnotesize
$\IP\left(\begin{matrix}1\,1\,1\,1\,1\,1\,0\cr 1\,1\,0\,0\,0\,0\,1\cr \end{matrix}\right)\hskip-3pt
	\left[\begin{matrix} 3~3\cr 3~0\cr \end{matrix}\right] \quotient{(\IZ_3:1,2,1,2,0,0,0,0)}$
&\cite{KreuzerRiegler}\\\hline
\str (0,20) & (10,10) & --                 &\cite[6.17]{GrossPopescu}\\\hline
\str (0,16) & (8,8)     &      --            &\cite[2.2]{GrossPopescu}\\\hline
\Str (-6,15) & (6,9)   &\footnotesize $\cicy{\IP^3\\ \IP^3\\}{1&3&0\\ 1&0&3\\}\quotient{A}$ & \SS1\\\hline
\str (0,14) & (7,7)     &  $X^{19,19}/\{\IZ_3,\,\IZ_2{\times}\IZ_2\}$  & \cite{BouchardDonagi}\\\hline
\Str (-18,13) & (2,11) &\footnotesize  $\cicy{\IP^2\\ \IP^2\\}{3\\ 3\\}\quotient{A{\times}D}$  & \SS1\\\hline
\str (0,12) & (6,6)     & --                 & \cite[4.10]{GrossPopescu}\\\hline
\SSTR (-16,10) & (1,9) &\footnotesize\hskip-3pt
$\renewcommand{\arraystretch}{0.8}
\cicy{\IP^1\\ \IP^1\\ \IP^1\\ \IP^1\\ \IP^1\\}{1~1\\ 1~1\\ 1~1\\ 1~1\\ 1~1\\}\quotient{\IZ_5}$\hskip-10pt,
$\IP_5[3,3]\quotient{\IZ_3{\times}\IZ_3}$,~ 
$\IP^7[2,2,2,2]/\{\IH,\IZ_4{\times}\IZ_2,\IZ_2{\times}\IZ_2{\times}\IZ_2\}$\hskip-3pt\null 
& \cite{Beauville, HjPark}\\\hline
\str (0,10) & (5,5)     & $X^{19,19}/\IZ_4$ & \cite{BouchardDonagi}\\\hline
\str (0,8) & (4,4)       & --                 & \cite[3.2]{GrossPopescu}\\\hline
\str (-8,6) & (1,5)      & $\IP^4[5]/\IZ_5{\times}\IZ_5 $ & --\\\hline
\str (0,6) & (3,3)  & $X^{19,19}/\{\IZ_3{\times}\IZ_3,\, \IZ_4{\times}\IZ_2,\, \IZ_3{\times}\IZ_2,\, \IZ_5\}$                &  \cite{BouchardDonagi}\\\hline
\str (0,4) & (2,2)       & -- (three families of manifolds) 
&\hskip-3pt{\small \cite[\!5.8, \!6.9, \!7.5]{GrossPopescu}}\hskip-3pt\null\\\hline
\end{tabular}
\parbox{6.5in}{\caption{\label{tiptab}\small 
The manifolds with $y=h^{11}{+}h^{21}\leq 22$ from 
\fref{tipplot}. In the `Manifold' column $X^{19,19}$ denotes the split bicubic and multiple quotient groups indicates different quotients with the same Hodge numbers. A dash indicates that the manifolds do not have simple names; all these manifolds arise~\cite{GrossPopescu} as special conifold transitions from well-known geometries. The $\IH$ in the quotient of $\IP^7[2,2,2,2]$ denotes the quaternion group. The vectors appended to the symmetries of the two weighted CICY's indicate how the generators act. The generator $(\IZ_3:1,2,1,2,0,0,0,0)$, for example, acts by multiplying the first coordinate by $\o$, the second by $\o^2$, etc., with $\o$ a nontrivial cube root of unity.
For each manifold with $\ch<0$ there is a mirror which we do not list explicitly.}}
\end{center}
\end{table}
In a recent article, Bouchard and Donagi \cite{BouchardDonagi} make a detailed classification of quotients of the split bicubic and find fundamental groups that are reproduced in the following table:
$$
\def\str{\vrule height14pt width0pt depth6pt}
\begin{tabular}{| c | c |}
\hline \str Group & $~(h^{11},\,h^{21})~$\\
\noalign{\hrule\vskip3pt}
\hline \str $~~\IZ_3\times\IZ_3~,~\IZ_4\times\IZ_2~,~\IZ_3\times\IZ_2~,~\IZ_5~~$ & $(3,\,3)$\\
\hline \str $\IZ_4$ & $(5,\,5)$\\
\hline \str $\IZ_2\times\IZ_2~,~\IZ_3$ & $(7,\,7)$\\
\hline \str $\IZ_2$ & $(11,\,11)$\\
\hline
\end{tabular}$$

To emphasize the paucity of manifolds with both Hodge numbers small we present in~\fref{tipplot} and~\tref{tiptab} the tip of the plot of Hodge numbers for $h^{11}+h^{21}\leq 40$ including the CICY's together with their mirrors, the toric CICY's, the toric conifolds, the quotient manifolds of which we are aware, a special class of manifolds fibered by abelian surfaces due to Gross and Popescu~\cite{GrossPopescu}, and certain interesting examples due to Tonoli~\cite{Tonoli}. For two of these manifolds, those with $\hodgenos=(1,23),\,(1,31)$, we do not show the points corresponding to the mirror manifolds since the constructions are such that the mirror manifolds are not known to exist. Our observation is that the tip is sparcely populated with some of the lowest points corresponding to manifolds we have discussed explicitly above or their quotients. The Kreuzer--Skarke list contains many pairs of points with $\ch=\pm 6$. These have Hodge numbers $(h,h+3)$ and $(h+3,h)$ for certain values of $h$ in the range $13\leq h\leq 128$ and it is easier to state the values of $h$ that are not found. These excluded values are 
$h=102, 103, 115, 117, 119\,\mbox{--}\,126$.
It is interesting that the Tian--Yau manifold, with $\hodgenos=(6,9)$ has Hodge numbers that are smaller than these other manifolds, which are simply connected.

We have made the observation that in order to find manifolds with low values of $h^{11}{+}h^{12}$ it is good to seek manifolds with a nontrivial fundamental group. The fundamental group, however, cannot be quite the right attribute since it is not respected by mirror symmetry. In \fref{tipplot}, for example, $\IP^4[5]/\IZ_5$ has fundamental group $\IZ_5$ while its mirror has trivial fundamental group. The attribute that we are after, for a manifold $Y$, is torsion in the integer cohomology ring $H^\bullet(Y,\IZ)$; for a clear discussion 
see~\cite{Batyrev:2005jc}. The torsion is a finite component of the cohomology ring that is absent if we work over $\IC$ or $\IR$. For the case that $H^1(Y,\IZ)\cong 0$ the cohomology groups take the general form
\begin{center}
\begin{tabular}{rlrl}
$H^0(Y,\IZ)$& $\cong~\IZ$ &$H^6(Y,\IZ)$ &$\cong~\IZ$  \\[5pt]
$H^1(Y,\IZ)$& $\cong~0$    &$H^5(Y,\IZ)$ &$\cong~{\mathfrak A}(Y)^*$  \\[5pt]
$H^2(Y,\IZ)$& $\cong~{\mathfrak A}(Y)\oplus\IZ^{h^{11}}$ &$H^4(Y,\IZ)$
   &$\cong~{\mathfrak B}(Y)^* \oplus\IZ^{h^{11}}\hskip-0.6in$ \\[10pt]
                   &$\hskip0.5in H^3(Y,\IZ)~\cong~{\mathfrak B}(Y)\oplus\IZ^{2h^{21}+2}\hskip-0.5in$\\
\end{tabular} 
\end{center}
where ${\mathfrak A}(Y)$ and ${\mathfrak B}(Y)$ are finite groups and  ${\mathfrak A}(Y)^*$  and ${\mathfrak B}(Y)^*$ are the corresponding dual groups. The group ${\mathfrak B}(Y)$ is the torsion of $H^3(Y,\IZ)$ and is known as the Brauer group. The group ${\mathfrak A}(Y)$ is closely related to the fundamental group through the isomorphism
$$
{\mathfrak A}(Y)\cong \text{Hom}(\p_1(Y),\, \IQ/\IZ)~.
$$ 
On the other hand, it is conjectured that under mirror symmetry, there is a relation
$$
{\mathfrak A}(Y^*)\oplus {\mathfrak B}(Y^*)^* \cong {\mathfrak A}(Y)^* \oplus{\mathfrak B}(Y)
$$ 
where $Y^*$ is the mirror of $Y$.

For the 16 examples of toric free group actions found by Batyrev and Kreuzer, it is the case that if a manifold $Y$ has a nontrivial fundamental group then its Brauer group is trivial and the mirror $Y^*$ has trivial fundamental group but nontrivial Brauer group. For manifolds defined by more than one polynomial, however, there are manifolds for which both ${\mathfrak A}$ and ${\mathfrak B}$ are simultaneously nontrivial. Thus the attribute that we are seeking is nontrivial torsion in the homology ring. Indeed, one of the manifolds~\cite[Thm. 6.9]{GrossPopescu} at the current tip of the cone with Hodge numbers $(2,2)$, a resolution of a very special nodal $\IP^7[2,2,2,2]$, is simply connected, but has recently been shown~\cite{GrossPavanelli} to have Brauer group ${\mathfrak B}(Y)\cong\IZ_8{\times}\IZ_8$, the largest known; its mirror~\cite[Rem. 1.5]{GrossPavanelli} is conjectured to have torsion in both its fundamental group and Brauer group.

Let us remark finally that string compactifications are often asymmetrical with respect to mirror symmetry. For both the models based on the Tian-Yau manifold and the quotient of the split bicubic, the Hosotani mechanism is used to reduce the spacetime gauge group, which requires a nontrivial fundamental group. It is compelling that there should be a mirror description of these models with the role of the fundamental group reflected onto the Brauer group. It would be of considerable interest to understand this relation.

\subsection{Key points}
This has been a long introduction, perhaps overly beset by detail, so let us summarize our three main points:
\begin{itemize}
\item The geography of \cys\ has a `tip' which appears to be sparsely populated. The sparse population seems to be a reflection of the fact that \cys\ whose homology has nontrivial torsion seem to be genuinely rare.
It is striking that the tip contains the manifold, that we shall call $X^{3,3}$, a quotient of  $X^{19,19}$ introduced in \eref{splitbicubic}, for which there is a heterotic model that has the particle content of the MSSM. The tip also contains the Tian--Yau manifold for which there is also a three generation model. The fact that the tip is sparsely populated makes the fact that we find two three-generation models here more surprising. The fact that $X^{3,3}$ is almost at the very end of the tip is a fact that would still be true even if new constructions increase the population.
\item It is natural to ask if the two three generation models are related and an answer is that the manifolds on which they are based are indeed closely related by conifold transitions. The most direct relation is that the Tian--Yau manifold is related via a conifold transition to a manifold $X^{7,7}$ which is a three-fold covering space of $X^{3,3}$.
\item It is natural also to ask, in relation to transitions between heterotic models, if it is possible to transfer bundles across a conifold transition. We refer to this process here as a transgression of bundles. A necessary condition for a bundle to arise as a transgression on the split manifold is that the bundle should be trivial on the lines that arise as the blow ups of the nodes. Remarkably this is the case for the heterotic bundle on $X^{3,3}$, suggesting that the heterotic bundle can be thought of as arising in this~way.
\end{itemize}

We have explained the first point in this introduction. In the remainder of the paper we elaborate on the second and third points. In \SS2 we discuss the relations between the bicubic and the split bicubic and their quotients. An interesting fact that is not obvious at the outset is that we may pass from the covering space of the Tian--Yau manifold to the split bicubic via a conifold transition
$$
\cicy{\IP^{3}\\ \IP^{3}\\}{1 & 3 & 0\\ 1 & 0 & 3\\}~\squig~~
\cicy{\IP^1\\ \IP^{2}\\ \IP^{2}\\}{1 & 1\\ 3 & 0\\ 0 & 3\\}\cicystop
$$
It follows, upon taking the quotient by $A\cong\IZ_3$, that the Tian--Yau manifold is related via a conifold transition to the quotient $X^{7,7}=X^{19,19}/A$ of the split bicubic, a manifold also in the tip with 
$\hodgenos=(7,7)$ and which is a three-fold cover of~$X^{3,3}$. 

Given that the the three-generation manifolds that we are considering are related by conifold transitions, we turn in \SS3 to the question of whether the vector bundles of their heterotic models are related. It is an old suggestion that it should, in certain circumstances, be possible to transfer bundles across a conifold transition. We examine this process. Although we have not been able to relate the tangent bundle of the Tian--Yau manifold to the vector bundle of $X^{3,3}$ directly, we note that it should be possible to transgress the tangent bundle of the Tian--Yau manifold to a, hitherto unknown, bundle on the $X^{7,7}$-manifold with $\ch=-6$. The fact that the vector bundle on $X^{3,3}$ is trivial on the `conifold lines', that is on the $\IP^1$'s that arise from the split
$$
\cicy{\IP^{2}\\ \IP^{2}\\}{3\\ 3\\}\quotient{A{\times} D}^{2,11}~\leftrightsquigarrow~~~~
\cicy{\IP^1\\ \IP^{2}\\ \IP^{2}\\}{1 & 1\\ 3 & 0\\ 0 & 3\\}\quotient{A{\times} D}^{3,3}
$$
suggests that the bundle can be thought of as arising from the manifold on the left. We examine this process and find, via a monad construction, some candidate bundles with the right Chern classes although we are not yet able to answer the question in the affirmative.

Seeing the tip of the landscape in \fref{tipplot} it is hard not to speculate on the possibility of a dynamical mechanism that would allow the universe to drift towards the tip. This of course is what makes the possibility of transgression so interesting. The burden of our discussion of transgression in \SS3 is that although the manifolds that we discuss seem to be discretely different, and the plots reinforce that impression, nevertheless the parameter spaces of different heterotic models meet in certain mildly singular manifolds and it is natural to ask if it is possible for the universe to move among these models. We conclude with a brief speculation along these lines in \SS4.
\subsection{Dramatis person\ae}
Here we list the principal actors of the story for reference and to fix notation for the rest of the paper. 

\begin{itemize}

\item In the Tian--Yau sequence, we have the three families of spaces
$$
(N'')^{14,23}~=~{\cicy{\IP^{3}\\ \IP^{3}\\}{1&3&0\\ 1&0&3\\}}~,~~
(N')^{8,35}~=~{\cicy{\IP^{2}\\ \IP^{3}\\}{3&0\\ 1&3\\}}~,~~
N^{\,2,83}~=~{\cicy{\IP^{2}\\ \IP^{2}\\}{3\\ 3\\}}~
$$
together with (resolutions of) respective quotients
$$
(M'')^{6,9}=(N'')^{14,23}/A~,~~~(M')^{6,9}=\widehat{(N')^{8,35}/A{\times }B} ~~
\mbox{and}~~
M^{6,9}=\widehat{N^{\,2,83}/A{\times} B{\times} C}.
$$
The actions of the groups $A$, $B$ and $C$ are explained 
in~\SS\ref{subsec_threethreegen}; the hats indicate resolutions of singularities of
the quotients. We will show that $M$, $M'$ and $M''$ all belong to the same
irreducible family and so following \SS\ref{subsec_threethreegen}, 
they will all be denoted by $M^{6,9}$.
\vskip10pt
\item The bicubic $N^{\,2,83}$ admit further free quotients
$$
N^{\,2,29}~=~\cicy{\IP^2\\ \IP^2\\}{3\\ 3\\}\quotient{A}~~~~\text{and}~~~~
N^{\,2,11}~=~\cicy{\IP^2\\ \IP^2\\}{3\\ 3\\}\quotient{A{\times}D}
$$
that are studied in~\SS\ref{subsec_quotientbicubic}.
\vskip10pt
\item In the split bicubic sequence, we have the family
\[
X^{19,\,19} ~=~\cicy{\IP^{1}\\ \IP^{2}\\ \IP^{2}\\}{1&1\\3&0\\ 0&3\\}
\]
as well as free quotients 
\[
X^{7,7}=X^{19,19}/A~,~~~\mbox{and}~~X^{3,3}=X^{19,19}/A{\times} D,
\]
where the actions of the groups $A$ and $D$ are explained 
in~\SS\ref{subsec_twoquotients}.

\end{itemize}
\newpage
\section{Relations Between Three-Generation Manifolds}
In this section we first explain the relation between the three equivalent presentations of the Tian--Yau manifold. Then we will examine the various conifold transitions between the bicubic and the split bicubic and their quotients.
\subsection{Three families of three--generation manifolds}
\label{subsec_threethreegen}
Consider the three families of manifolds
$$
(N'')^{14,23}~=~\Mone^{14,23}~,~~
(N')^{8,35}~=~\Mtwo^{8,35}~,~~
N^{\,2,83}~=~\Mthree^{2,83}~,
$$
as well as resolutions of certain quotients
$$
(M'')^{6,9}=(N'')^{14,23}/A~,~~~(M')^{6,9}=\widehat{(N')^{8,35}/A{\times }B} ~~
\mbox{and}~~
M^{6,9}=\widehat{N^{\,2,83}/A{\times} B{\times} C}.
$$
Here the groups $A,\ B$ and $C$ are all abstractly isomorphic to $\IZ_{3}$. The group $A$ acts freely, while $B$ and $C$ have fixed curves that are tori. The upshot is that the Euler numbers $\ch=2(h^{11}{-}h^{21})$ of the (resolved) quotient manifolds are obtained from that of the covering manifold by dividing by the orders of the groups, with the result that $M$, $M'$ and $M''$ all have $\chi=-6$. We will see also that the three manifolds also all have Hodge numbers $\hodgenos=(6,9)$. As stated previously, these families of manifolds have been shown to belong to the same irreducible component of the moduli space in~\cite{GreeneKirklin}, but it is useful to review some aspects of this correspondence in order to establish conventions and notation. A detailed study of the Tian--Yau manifold~$(M'')^{6,9}$, and of a three-generation model based on this manifold, is to be found 
in~\cite{GreeneKirklinMironRoss}.

We begin by discussing the Tian--Yau manifold. The covering space~$(N'')^{14,23}$ is described by three equations of the indicated bidegrees. One may take, for example, the equations
\beq
\begin{split}
F&= f_{0}\, x_{0}y_{0} + f_{1}\sum_j  x_{j}y_{j}
+ f_{2}\sum_j  x_{j}y_{j+1} +  f_{3}\sum_j  x_{j+1}y_{j}
+ f_{4}x_{0}\sum_j  y_{j} +  f_{5}(\sum_j  x_{j})y_{0}~,\\
G&= x_{0}^{3} -  x_{1}x_{2}x_{3} + g_1\sum_j  x_{j}^{3} + 
g_{2}\, x_{0}\sum_j  x_{j}x_{j+1}~,\\
H&= y_{0}^{3} - y_{1}y_{2}y_{3} + h_1\sum_j  y_{j}^{3} + 
h_{2}\, y_{0}\sum_j  y_{j}y_{j+1}~.
\end{split}
\label{FGH}\eeq
where the $(x_{0},x_{j})$ and $(y_{0},y_{j})$, $j=1,2,3$, are projective coordinates for the two                  $\IP^{3}$'s and $f_a$, $g_a$ and $h_a$ are coefficients. The separate treatment of the zeroth coordinate anticipates the action of the symmetry group  $A\cong\IZ_{3}$
with generator
\beq
A:~~(x_{0},x_{j}){\times}(y_{0},y_{k})  \mapsto (x_{0},x_{j+1}){\times}(y_{0},y_{k+1})~,
\label{Agen}\eeq
with the $j$ and $k$ indices understood as being reduced mod 3.
The freedom to make changes of coordinates has been used to ensure that in $G$ the term 
$x_1 x_2 x_3$ appears with coefficient~$-1$ and the terms of the form 
$x_{\mu}^{2}x_{\nu}$, with $\m\neq\n$, appear with coefficient zero, and similarly for~$H$. The polynomial $F$ may be redefined by an overall scale, but apart from this all the coefficients are significant. Apart from this freedom to make coordinate changes and scale $F$, these are the most general polynomials invariant under $A$, yielding a total of $10{-}1=9$ parameters.
To check that $A$ acts on $(M'')^{6,9}$ without fixed points, note that
the fixed point set of $A$ in the first $\IP^3$ consists of the curve
$(x_{0}, x_{1},x_{1},x_{1})$ and two points of the form 
$(0,1,\o,\o^{2})$ with $\o$ a nontrivial cube root of unity. The two isolated points do not satisfy the equation $G=0$, for general values of the parameters, and so do not lie on $(M'')^{6,9}$. It remains to check the points of the form
 $$
 (x_{0}, x_{1},x_{1},x_{1})\times (y_{0}, y_{1},y_{1},y_{1})~. $$
and it is easy to see that these points do not satisfy the equations for general values of the parameters. Thus the quotient is smooth and has Euler number -6. The Hodge numbers may be calculated individually by applying the Lefschetz hyperplane theorem; however, a counting of the number of ways in which the equations may be deformed works in this case and gives $h^{21}=9$, so $h^{11}=6$.

Next we turn to 
$$
(N')^{8,35}~=~\Mtwo $$
and write $(\xi_{1},\xi_{2},\xi_{3})$ for the projective coordinates of the $\IP^{2}$ and, as before, 
$(y_{0},y_{j})$ for the coordinates of the $\IP^3$. We write the equations for this space in the form
\beq\begin{split}
F&= f_{0}\, x_{0}y_{0} + f_{1}\, \sum_{j}x_{j}y_{j}
+  f_{2}\, \sum_{j}x_{j}y_{j+1} +  f_{3}\, \sum _{j}x_{j+1}y_{j}
+ f_{4}\, x_{0}\sum_j y_{j} +  f_{5}\, (\sum_j x_{j})y_{0} ~,\\
H&= y_{0}^{3} - y_{1}y_{2}y_{3} + h_1\,\sum_{j}y_{j}^{3} + 
h_{2}\, y_{0}\sum_{j}y_{j}y_{j+1}~.\notag
\end{split}
\eeq
The polynomials $F$ and $H$ are as before but we take now 
\beq
x_{0}=\xi_{1}\xi_{2}\xi_{3}~,~~x_{j}=\xi_{j}^{3}~,~j=1,2,3.
\label{x-xi}\eeq
Note that with this identification the $x$'s satisfy the singular cubic 
$$
G_0\defineas x_{0}^{3} - x_{1}x_{2}x_{3} = 0~.$$
We have a symmetry $A$ that acts cyclically as before and also a new symmetry, $B$, that acts on the $\xi_{j}$ leaving the $x$'s and $y$'s unchanged:
\beq
\begin{split}
A:\hskip0.1in \xi_{j}&\to \xi_{j+1}~,~~(y_{0},y_{j}) \mapsto (y_{0},y_{j+1})~,\\
B:\hskip0.1in \xi_{j}&\to \o^{j}\xi_{j}~.
\end{split}\notag
\eeq
Note that one consequence of identifying the $\xi$'s under $B$ is to render the map $\xi\to x$ injective.

The symmetry $A$ acts without fixed points as before but $B$ has fixed points in its action on 
$\xi$ which are $\{(1,0,0),\,(0,1,0),\,(0,0,1)\}$. These lead, via $F$, to a linear constraint on the $y$'s which, in conjunction with the equation $H=0$, leads to three fixed tori which are identified under the action of   $A$. Since the fixed point set has Euler number zero the Euler number of the singular quotient is ${-}6$ and the Euler number of the resolved quotient is also ${-}6$ since the resolution replaces the singular torus by a torus multiplied pointwise by the $A_{2}$ Eguchi--Hanson surface. The parameter count proceeds by noting that we now have $2+5=7$ parameters in the equations and the $A_{2}$ Eguchi--Hanson surface has two (1,1)-forms from which one may form two (2,1)-forms on the threefold by multiplying these by the holomorphic form of the torus. Thus we have $h^{21}=7+2=9$ and hence also $h^{11}=6$ as~before.

In fact, it is easy to see from our discussion that a resolved quotient
$(M')^{6,9}$ deforms back smoothly to a Tian--Yau manifold $(M'')^{6,9}$.
After taking the quotient by $A\times B$, the space is given by the
equations $F, G_0$ and $H$, and contains a curve of $A_2$ singularities
along the torus discussed above. To obtain $(M')^{6,9}$, we resolve this
singularity; to obtain $(M'')^{6,9}$, we smoothen the defining equation
$G_0$ to a smooth $G$. However, it is well known that the resolution and the
deformation of a surface $A_2$ singularity are smooth deformation
equivalent, and this also works for the family of singularities parametrized
by the torus. Hence indeed, $(M')^{6,9}$ can be deformed to $(M'')^{6,9}$.

For the third manifold 
$$
N^{\,2,83}~=~\Mthree $$
the construction is performed twice taking the equation to be
$$
F = f_{0}\, x_{0}y_{0} + f_{1}\sum_j x_{j}y_{j}
+  f_{2}\sum_j x_{j}y_{j+1} +  f_{3}\sum_{j}x_{j+1}y_{j}
+ f_{4}\, x_{0}\sum_j y_{j} +  f_{5}(\sum_j x_{j})y_{0}~,
$$
where now
$$
x_{0}=\xi_{1}\xi_{2}\xi_{3}~,~~x_{j}=\xi_{j}^{3}~,~j=1,2,3,
~~\hbox{and}~~~
y_{0}=\eta_{1}\eta_{2}\eta_{3}~,~~y_{j}=\eta_{j}^{3}~,~j=1,2,3.
$$
We now have symmetries $A$ and $B$ as before and a new symmetry $C$, for which it is convenient to take the generator to be 
$$
C~:~~\eta_{j}\mapsto \o^{-j}\eta_{j}~.
$$ 
The symmetry $A$ acts without fixed points while $B$ and $C$ each have a fixed torus.
The equation $F$ is transverse as a function of the $\xi$'s and $\eta$'s for general values of the coefficients. A transverse $F$ is obtained by taking, for example,
$f_{0}=f_{1}=1$, $f_{2}=f_{3}=c$ and $f_{4}=f_{5}=0$ providing 
$c\neq 0,1$. The parameter count is again $h^{21}=5+2+2=9$ with 5 parameters coming from $F$ and two each from the resolution of the two fixed tori. The same argument as before shows that $M^{6,9}$ also deforms smoothly to
$(M')^{6,9}$, and hence to $(M'')^{6,9}$.

We have shown that $M$, $M'$ and $M''$ all have Hodge numbers
$(6, 9)$, and that these three manifolds belong to the same irreducible
family. From now on we will omit the primes since no confusion can arise
and will choose at various times whichever presentation is convenient.
\subsection{Some quotients of the bicubic} 
\label{subsec_quotientbicubic}
We now examine briefly two related quotients of the manifold $N^{\,2,83}$. First, there is 
$$
N^{2,29}\defineas~~\cicy{\IP^2\\ \IP^2\\}{3\\ 3\\}\quotient{A}, 
$$
a free quotient with Euler number $\ch=-54$. Next, we also have
$$
N^{2,11}\defineas~~\cicy{\IP^2\\ \IP^2\\}{3\\ 3\\}\quotient{A{\times}D}
$$
that we will need in the following, where we denote by $D$ the diagonal subgroup of $B{\times}C$ with generator 
$$
D~:~~\x_j \to \o^j \x_j~,~~~\eta_k\to \o^{-k}\eta_k~.
$$
There are a total of 12 independent polynomials invariant under $A{\times D}$ that are given in the following table:
$$
\def\str{\vrule height16pt width0pt depth8pt}
\begin{tabular}{| l | c | c | c | c | c | c |}
\hline 
\str Polynomial 
&$\sum \x_j^2\x_{j\pm 1}\eta_{j+k}^2\eta_{j+k\pm 1}$
   &$\sum \x_j^3\eta_{j+k}^3$ &$\x_1\x_2\x_3\sum\eta_j^3$ 
      & $\left(\sum \x_j^3\right)\eta_1\eta_2\eta_3$ 
          &$\x_1\x_2\x_3 \eta_1\eta_2\eta_3$\\
\hline
\str Number&6&3&1&1&1\\
\hline
\end{tabular}$$
It is easy to see that $A{\times}D$ acts without fixed points for generic coefficients of the defining polynomial. Since there are 12 coefficients and an overall scale is irrelevant we have $h^{21}=11$ and since the group action is fixed point free we have $\ch=-18$ so $h^{11}=2$.
\subsection{Two quotients of the split bicubic}
\label{subsec_twoquotients}
Having addressed the three avatars of the Tian--Yau manifold that were central to the discussion of three generation models in early investigations, we turn now to the split bicubic $X^{19,19}$ and its quotients 
$$
X^{7,7}~=~~\cicy{\IP^1\\ \IP^2\\ \IP^2}{1&1\\ 3&0\\ 0&3\\}^{7,7}\quotient{A}~~~\mbox{and}~~~~~~
X^{3,3}~= ~~\cicy{\IP^1\\ \IP^2\\ \IP^2}{1&1\\ 3&0\\ 0&3\\}^{3,3}\quotient{A{\times}D}\cicystop
$$
For these manifolds take coordinates $t_{r}$, $r=1,2$ on the $\IP^1$, and coordinates $\xi_{j}$ and $\eta_{j}$ on the two $\IP^2$'s as previously. The generator of $A$ can be chosen to act cyclically as before leaving the $t$'s invariant~\cite{Braun:2007tp}:
 $$
A:\hskip0.1in \xi_{j}\to\xi_{j+1}~,~~\eta_{j}\to\eta_{j+1}~,~~t_{r} \to t_{r}~.$$
The group $D$ is the diagonal subgroup of $B{\times}C$ considered above and which we also take to leave the coordinates of the $\IP^1$ invariant:
 $$
D:\hskip0.1in \xi_{j}\to\o^{j}\xi_{j}~,~~\eta_{j}\to\o^{-j}\eta_{j}~,~~t_{r} \to t_{r}~.$$
Under $A{\times}D$ the only cubic monomials in the $\x_j$ that are invariant are $\sum_j\xi_{j}^{3}$ and 
$\xi_{1}\xi_{2}\xi_{3}$ and similarly for the $\eta_k$. With a certain choice of $t$ coordinate we may write the polynomials for the $A{\times}D$ quotient in the form
\beq
\begin{split}
F^1&=~t_1\Big(\sum_{j=1}^3\x_j^3 -3a\,\x_1\x_2\x_3\Big)+ 3c t_2\,\x_1\x_2\x_3 \\[3pt]
F^2&=~3c t_1\,\eta_1\eta_2\eta_3  + t_2\Big(\sum_{j=1}^3\eta_j^3 - 3b\,\eta_1\eta_2\eta_3\Big)
\label{ADcubics}
\end{split}
\eeq
and the coefficients $a,b,c$ give the correct count for $h^{21}(X^{3,3})=3$.
Under~$A$ the fixed point set, in the embedding space, is of the form
$$
(t_1,t_2)\times(1,\a,\a^2)\times(1,\b,\b^2)~;~~~\a^3=\b^3=1~.$$
and these points do not satisfy the equations unless $c^2=(a-1)(b-1)$. The fixed point set under~$D$ is
 $$
 (t_{1},t_{2})\times\Big\{(1,0,0),\,(0,1,0),\,(0,0,1)\Big\}\times
 \Big\{(1,0,0),\,(0,1,0),\,(0,0,1)\Big\}$$ 
and these points do not satisfy the equations for $(t_1,t_2)$ a point in~$\IP^1$. 

If the equations are merely to be invariant under~$A$, then we are allowed the terms $\sum_j \x_j^2\x_{j\pm 1}$ and $\sum_j \eta_j^2\eta_{j\pm 1}$. We are also now free to make coordinate redefinitions $\x_j\to \x_j+r\x_{j+1}+s\x_{j-1}$ and similarly for $\eta$. We may use this freedom to bring the equations to the form
\beq
\begin{split}
F^1&=~t_1\Big(\sum_{j=1}^3\x_j^3 - 3a\,\x_1\x_2\x_3 + 
d_{+}\sum_{j=1}^3\x_j^2\x_{j+1} + d_{-}\sum_{j=1}^3\x_j^2\x_{j-1}\Big) + 
3ct_2\,\x_1\x_2\x_3 \\[3pt]
F^2&=~3ct_1\,\eta_1\eta_2\eta_3 + 
t_2\Big(\sum_{j=1}^3\eta_j^3 - 3b\,\eta_1\eta_2\eta_3 + 
e_{+}\sum_{j=1}^3\eta_j^2\eta_{j+1} + e_{-}\sum_{j=1}^3\eta_j^2\eta_{j-1}\Big)
\label{Feqs}
\end{split}
\eeq
which exhibit the seven complex structure parameters corresponding to $h^{21}(X^{7,7})=7$.
\newpage
\subsection{Conifold transitions between the manifolds}
\newcommand{\bc}{\cicy{\IP^2\\ \IP^2\\}{3\\3\\}}
\newcommand{\sbc}{\cicy{\IP^1\\ \IP^2\\ \IP^2}{1&1\\3&0\\0&3\\}}
\newcommand{\lrsquig}{\leftrightsquigarrow}
\newcommand{\darr}{\downarrow}
\begin{table}[!ht]
\begin{center}
\boxed{
\begin{tabular}{rlcll}
\\
&$\bc^{2,83}$ &  $\hskip-5pt\lrsquig\hskip15pt$  &$\sbc^{19,19}$ \\
&$\hskip 18pt\darr$&&$\hskip28pt\darr$\\
&$\bc^{2,29}\quotient{A}$ & $\hskip-5pt\lrsquig\hskip15pt$  &$\sbc^{7,7}\quotient{A}$ \\
&$\hskip 18pt\darr$&&$\hskip28pt\darr$\\
&$\bc^{2,11}\quotient{A{\times}D}$ &  $\hskip-5pt\lrsquig\hskip15pt$ 
  &$\sbc^{3,3}\quotient{A{\times}D}$ \\
&$\hskip 18pt\downharpoonright$&&$\hskip28pt \downharpoonright $\\
$\hskip11pt\cicy{\IP^3\\ \IP^3\\}{1&3&0\\ 1&0&3\\}\quotient{A}^{6,9}~\cong$
     &$\widehat{\bc}\quotient{A{\times}B{\times}C}^{6,9}$ 
        & $\hskip-5pt\lrsquig\hskip15pt$ 
           &$\widehat{\sbc}^{7,7}\quotient{A{\times}B{\times}C}$
               &$\hskip-10pt\cong~~~~\sbc\quotient{A }^{7,7}\hskip11pt$\\
\\              
\end{tabular}}
\vskip5pt
\parbox{5.5in}{\caption{\label{transitions}\small Birational relations between quotients of the bicubic, split bicubic and the Tian--Yau manifold; hats denote resolutions of quotients.}}
\end{center}
\end{table}
\renewcommand{\kbrowstyle}{\relax}
\renewcommand{\kbcolstyle}{\relax}
\setlength{\kbrowsep}{7pt}
\setlength{\kbcolsep}{-2pt}
We have introduced two sequences of manifolds: (i) the Tian--Yau sequence 
$$
N^{\,2,83}\to N^{2,29}\to N^{2,11}~,
$$ 
of interest since the early days of string phenomenology and   (ii) the split-bicubic sequence 
$$
X^{19,19}\to X^{7,7} \to X^{3,3}~,
$$ 
recently of interest in constructing the MSSM. In this section, we show the remarkable relation that the two are related by a conifold transition, and moreover, in the next section, how bundles on one may be transferred to another by the process of transgression. \tref{transitions} below shows the intricate web of relations between the various manifolds we have been examining. These show the birational relations between the various quotients of the bicubic, split bicubic and the Tian--Yau manifold. The vertical arrows denote quotients by freely acting groups while the harpoons above the last row denote the process of taking a quotient by $C$ and then resolving the singularities. The relation expressed by the $\leftrightsquigarrow$ is that of a conifold transition~\cite{hubsch}.

Let us first briefly recall how a conifold transition works in the context of CICY's. This is the `splitting' process of which a good account may be found in~\cite{hubsch}. Let us take as an example the top row of 
\tref{transitions}. To the right is the split bicubic $X^{19,19}$. As in~\SS2.3, we can write its defining equations as
\beq\label{defX1919}
t_1 U(\xi) + t_2 W(\xi) = 0 \ , \qquad
t_1 Z(\eta) + t_2 V(\eta) = 0 \ ,
\eeq
where as before $t_{1,2}$ are the coordinates of the $\IP^1$,
$\xi_j, \eta_j$ are respectively the coordinates of the two
$\IP^2$'s, and $U,V,W,Z$ are cubic polynomials. We can regard
\eref{defX1919} as a
matrix equation in $t_{1,2}$. Now, because $t_{1,2}$ are projective
coordinates on $\IP^1$ they cannot vanish simultaneously, hence
\eref{defX1919} can only hold if
\beq
F_0=\det \left|
\begin{array}{cc}U(\xi) & W(\xi) \\
                           Z(\eta) & V(\eta)
\end{array} \right| = U(\xi)  V(\eta) - W(\xi)  Z(\eta) = 0 \ .
\label{Fsharp}\eeq
This is, a bicubic equation in the $\xi$ and $\eta$ coordinates. In  fact it is a singular limit of the manifold 
$N^{\,2,83}$. The limit is singular owing to the fact that all the derivatives of $F_0$ vanish precisely at the points where $U,V,W,Z$ are all zero and this happens, in this case, in $3^4$ points. We have seen that a point 
$(t,\x,\eta)$ that satisfies \eref{defX1919} must be such that $(\x,\eta)$ satisfy \eref{Fsharp}. Conversely suppose that $(\x,\eta)$ satisfy \eref{Fsharp} and that for these values the cubics $U,V,W,Z$ are not all zero. The equations \eref{defX1919} will then determine a unique ratio $t_1/t_2$ hence a unique point $t\in\IP^1$.
If, however, all four of the cubics vanish then the equations \eref{defX1919} are satisfied for all values of 
$t\in\IP^1$. For suitable cubics the split manifold defined by \eref{defX1919} is smooth while the conifold defined by $F_0=0$ is singular at a certain number of nodes. The split manifold projects down onto the conifold such that a unique point projects to every nonsingular point of the conifold but an entire 
$\IP^1$ of the split manifold projects down onto each node. Alternatively we pass from the conifold to the split manifold by blowing up each node to a $\IP^1$.

The first three rows of \tref{transitions} are simply conifold transitions between 
$N^{\,2,83}$ and    $X^{19,19}$ as well as their quotients. The relation between the extreme ends of the last row is less obvious and corresponds to the relation

\beq
\cicy{\IP^3\\ \IP^3\\}{1&3&0\\ 1&0&3\\}\quotient{A}^{6,9}\leftrightsquigarrow~~
\cicy{\IP^1\\ \IP^3\\ \IP^3}{1&1&0&0\\ 1&0&3&0\\ 0&1&0&3\\}\quotient{A}^{7,7}\cong~~
\cicy{\IP^1\\ \IP^2\\ \IP^2}{1&1\\ 3&0\\ 0&3\\}\quotient{A}^{7,7}
\label{TY-splitbicubic}\eeq
which we explain in the following. The relation is true whether or not we take the $A$-quotient, so we shall first explain the relation before taking the quotient. The isomorphism between the second and third families in the above relation may be surprising; it is this that we will now explain.

\subsubsection{Three avatars of the split bicubic}
Consider the family of complete intersection Calabi--Yau manifolds

\beq
\def\ns{\hskip-4pt}
\def\skip{\hskip6pt}
\kbordermatrix{&\ns\widetilde{G}^0\ns\!&\ns\widetilde{H}^0\ns\ns\skip&\ns G^1\ns&\ns G^2\ns
&\ns G^3\ns\skip&\ns\! H^1\ns\!&\ns\! H^2\ns\!&\ns\! H^3\ns\!\\
                        t~~~\IP^1&1&1\skip&0&0&0\skip&0&0&0\\
                      \x~~~\IP^2&0&0\skip&1&1&1\skip&0&0&0\\
                   \eta~~~\IP^2&0&0\skip&0&0&0\skip&1&1&1\\
                       x~~~\IP^3&1&0\skip&1&1&1\skip&0&0&0\\
                       y~~~\IP^3&0&1\skip&0&0&0\skip&1&1&1}~=~X~,
\label{extended}\eeq

where we have indicated the coordinates of the projective spaces and have also named the polynomials whose degrees correspond to the columns of the matrix. We will soon see that calling this manifold $X$ is justified since we shall show that $X/A$ is isomorphic to both the second and third CICY's in~\eref{TY-splitbicubic}. This is a matter of showing, firstly, that we obtain maps between the various manifolds by  eliminating variables from the defining equations and, secondly, that these maps are isomorphisms rather than being merely birational equivalences. 

Consider the three equations $G^j$. These are bilinear in the coordinates $\x$ and $x$ of the first $\IP^2$ and first $\IP^3$
$$
G^j~=~G^j{}_{k\a}\,\x^k x^\a~=~G(x)^j{}_k\, \x^k~;~~~~G(x)^j{}_k=G^j{}_{k\a}\, x^\a~,
$$
where we are using a summation convention and $j,k=1,2,3$ while $\a=0,..,3$. Now,
the equations 
$G(x)^j{}_k\, \x^k=0$ are three equations in the three $\x^k$ and these coordinates are not all zero. So we have
$$
G(x)~\defineas~\det\Big(G(x)^j{}_k\Big)~=~0~.
$$ 
We see that $G(x)$ is the determinant of a $3{\times}3$ matrix of linear forms and so is a cubic. It is also a classical fact that any cubic in variables $x^\a$, $\a=0,..,3$ can be written as the determinant of a matrix of linear forms (for a review of this and related facts see \cite{BuckleyKosir}; the original reference is 
\cite{Dixon1902}). Analogous considerations apply also to the equations $H^j$ leading to a determinant $H(y)$.

We have now reduced the triple $G^j$ (and respectively $H^j$) into a
single equation $G$ (and respectively $H$). It thus remains to the
show that
\beq
\def\ns{\hskip-3pt}
\kbordermatrix{&\ns\!\widetilde{G}^0\ns\!&\!\ns\widetilde{H}^0\ns\ns&\ns  G\ns&\ns H\ns\\
       t~~~\IP^1&1&1&0&0\\
      x~~~\IP^3&1&0&3&0\\
      y~~~\IP^3&0&1&0&3}
\label{intermediateX}\eeq
is another avatar of $X$. This is done by simply remarking that the rank of the matrix $G(x)^j{}_k$ is $\leq2$ owing to the fact that it has zero determinant. For a general point $x$ the rank is 2 and in fact this is true for all $x$ since to have rank 1 would impose 4 independent linear relations on $x\in \IP^3$ which cannot be satisfied in a three dimensional space. This being so there is a one-one relation between allowed points $\x$ and allowed points $x$ and the manifold that results from eliminating $\x$ and $\eta$ in this way is smooth. This establishes that the CICY~\eref{intermediateX} is isomorphic to $X$.

Finally, we show that $X$ is indeed the familiar split bicubic, $X^{19,19}$. This is done by eliminating variables in another way. The four equations $\widetilde{G}^0=0$ and $G^j=0$ can be written
$$
\widetilde{G}(t,\x)^\a{}_\b\, x^\b~=~0~~~\mbox{where}~~~
\widetilde{G}(\x)^j{}_\b=G^j{}_{k\b}\,\x^k~.
$$
Thus we have a set of four equations for the four coordinates $x^\b$, which cannot all be zero. It follows that
$$
F^1~\defineas~\det\Big(\widetilde{G}(t,\x)^\a{}_\b\Big)~=~0.
$$
Since $\widetilde{G}(t,\x)^0{}_\b$ is independent of $\x$ and the $\widetilde{G}(t,\x)^j{}_\b$ are independent of $t$ we see that $F^1$ has bidegree $(1,3)$ in $t$ and $\x$. Analogous considerations apply also to the matrix $\widetilde{H}(t,\eta)^\a{}_\b$ and~to 
$$
F^2~\defineas~\det\Big(\widetilde{H}(t,\eta)^\a{}_\b\Big)~.
$$
One checks, analogously to the previous case, that the matrices $\widetilde{G}(t,\x)^\a{}_\b$ and 
$\widetilde{H}(t,\eta)^\a{}_\b$ always have rank 3. We have now shown that indeed the manifold
$$
\def\ns{\hskip-7pt}
\kbordermatrix{&\ns F^1\ns&\ns F^2\ns\ns\\
       t~~~\IP^1&1&1\\
     \x~~~\IP^2&3&0\\
  \eta~~~\IP^2&0&3}~=~X~.
$$

\subsubsection{The conifold transition between $M$ and $X^{7,7}$}
We have now seen three equivalent representations of $X^{19,19}$.
Next, we proceed to show the relation proposed in
\eref{TY-splitbicubic}. This is seen easily since the determinants~$G$
and~$H$ become the cubics of the covering space of the Tian--Yau manifold
$$
\def\ns{\hskip-3pt}
\kbordermatrix{&\ns F\ns&\ns G\ns&\ns H\ns\\
      x~~~\IP^3&1&3&0\\
      y~~~\IP^3&1&0&3}
$$ 

On making the conifold transition it is possible to approach the conifold from the Tian--Yau side by singularizing the bilinear equation $F$ while leaving $G$ and $H$ fixed.  The coordinates $t$ can be chosen such that the equations $\widetilde{G}^0$ and $\widetilde{H}^0$ take the form
\bean
\widetilde{G}^0&=~t_1\left(\sum_j x_j - 3ax_0\right) + 3ct_2\,x_0\\
\widetilde{H}^0&=~3ct_1\,y_0 + t_2\left(\sum_j y_j - 3by_0\right)~.
\eean
Again we have two equations in the coordinates $(t_1,t_2)$, which cannot both be zero, so the determinant must vanish
\beq
F_0~\defineas~\left(\sum_j x_j - 3ax_0\right) \left(\sum_j y_j - 3by_0\right) - 9c^2 x_0y_0~=~0
\label{conifoldF}\eeq
and the polynomial $F$ is a deformation of $F_0$. Comparing with \eref{FGH} we see that the parameter count is that there are two parameters in each of $G$ and $H$ and three in $F_0$. This accounts for the 7 complex structure parameters of the split bicubic.

Consider now the manifold 
$$
\cicy{\IP^1\\ \IP^2\\ \IP^2\\}{1&1\\ 3&0\\ 0&3\\}^{3,3}\quotient{A{\times}D}
$$
One might at first seek to relate this to the Tian--Yau manifold by taking a quotient of~\eref{TY-splitbicubic} by 
$D$ but this cannot be done, owing to the fact that $D$ does not act on the Tian--Yau manifold. 
Another reason why this fails is that it is not possible to find a nonsingular CICY of the form \eref{extended} that is invariant under $A{\times}D$.

We have, instead, the splitting
\beq
\cicy{\IP^2\\ \IP^2\\}{3\\ 3\\}^{2,11}\quotient{A{\times}D} \leftrightsquigarrow~~~
\cicy{\IP^1\\ \IP^2\\ \IP^2\\}{1&1\\ 3&0\\ 0&3\\}^{3,3}\quotient{A{\times}D}
\label{threethreeid}\eeq
as in \tref{transitions} and we have seen previously that the group $A{\times}D$ acts freely on both manifolds. 
We can take the quotient by $C$ and then resolve. This gives us the Tian--Yau manifold on the left but the relation that we find is
$$
\cicy{\IP^3\\ \IP^3\\}{1&3&0\\ 1&0&3\\}\quotient{A}^{6,9} \leftrightsquigarrow~~~
\widehat{\cicy{\IP^1\\ \IP^2\\ \IP^2\\}{1&1\\ 3&0\\ 0&3\\}}\quotient{A{\times}B{\times}C}^{7,7} =~~~~~~~~
\cicy{\IP^1\\ \IP^2\\ \IP^2\\}{1&1\\ 3&0\\ 0&3\\}\quotient{A}^{7,7}
$$
where it should be noted that $B$ and $C$ have fixed points whose resolution is implicit in this identity. We also know from the previous subsection that
$$
\cicy{\IP^3\\ \IP^3\\}{1&3&0\\ 1&0&3\\}\quotient{A}^{6,9}
\leftrightsquigarrow~~~
\cicy{\IP^1\\ \IP^2\\ \IP^2\\}{1&1\\ 3&0\\
0&3\\}\quotient{A}^{7,7}\cicystop
$$
The upshot is that we again relate the Tian--Yau manifold to $X^{7,7}$ rather than to~$X^{3,3}$.

\newpage
\section{Transgression of Vector Bundles}
Given the close relation between the Tian--Yau manifold and the quotient of the split bicubic, it is natural to ask how the tangent bundle of the Tian--Yau manifold is related to the vector bundles of the heterotic models based on the split bicubic.  We will describe here a process of transferring a bundle between two manifolds that are related by a conifold transition. Every manifold comes equipped with a tangent bundle, $\cT$, and a conifold transition  induces a change in the tangent bundles of the manifolds. This change is rather drastic since $c_3(\cT)$ is the Euler number of the manifold and, as the result of a conifold transition, this changes by twice the number of nodes of the conifold. This jump is the inevitable consequence of the fact that the tangent bundle is singular where the manifold fails to have a well defined tangent space. It was pointed out to one of the present authors, 
by Witten~\cite{WittenPrivate}, many years ago,  that other bundles are better behaved and can be expected to deform smoothly across the transition. This fits in well with heterotic models since there is no reason, in general, to prefer the tangent bundle over others. We will here apply this idea to the three-generation models that we have been discussing. As a matter of language we prefer to refer to the process of transferring bundles across a conifold transition as a
transgression\footnote{{\it transgress\/} from L.\ {\it transgredior} = {\it trans\/} + {\it egredior}, to go across, to cross over.} of bundles.
\vfil
\begin{figure}[!h]
  \begin{center}
\includegraphics[width=4.7in]{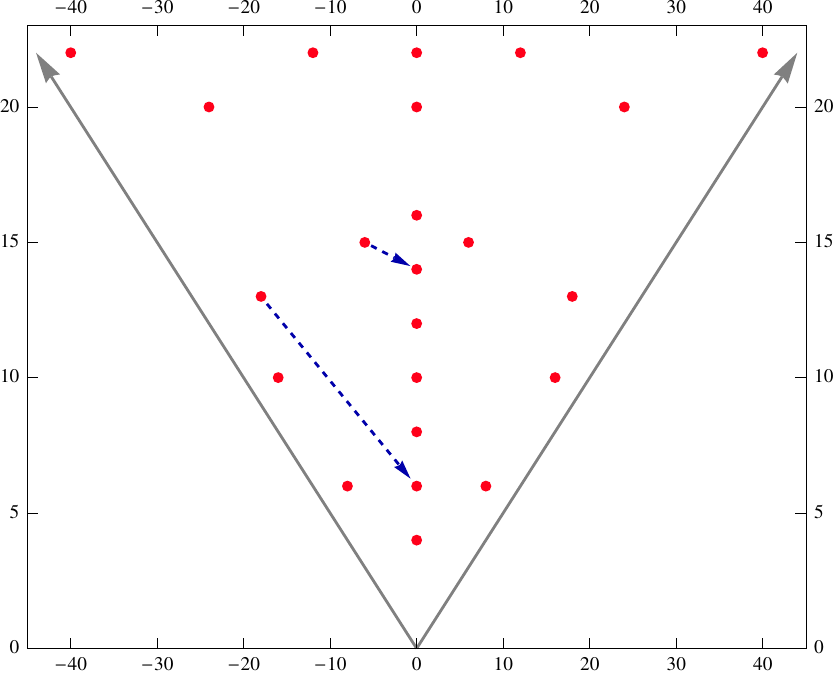}
\vskip5pt
\parbox{5.7in}{\caption{\label{transgressions}
\small The manifolds with $h^{11}+h^{21}\leq 22$ showing the two transgressions that we propose here between the Tian-Yau manifold and $X^{7,7}$, and between $N^{\,2,11}$ and $X^{3,3}$.}}
\end{center}
\end{figure}
\subsection{Local analysis of transgressions}
In coordinates $w^j=(u,v,w,z)$ on $\IC^4$, a conifold singularity at the origin in $\IC^4$ is described locally by the vanishing of the equation
$$f_0=uv-wz~.$$
The conifold is smoothed to a manifold, that we denote by $M_t$, by deforming
$f_0$ to $f_t~=~f_0 + t g$ where $g$ is any function such that the derivatives 
of~$f_t$ can no longer all vanish simultaneously. The split manifold,
$\Mcheck$, is realized in $\IC^4{\times}\IP^1$ by the equations
\beq
\begin{split}
t_1 u + t_2 w~&=~0\\
t_1 z + t_2 v~&=~0~.
\end{split}
\eeq
The sections of the tangent bundle~$\cT_t$ of~$M_t$ consist of the set 
of vector fields~$V^j$ that are tangent to~$M_t$; this condition can be written as
\beq
V^j\pd{f_t}{w^j}~=~0.
\label{Tflat}\eeq
For $t\neq 0$, this condition ensures that $\cT_t$ has rank~3. However, at the conifold point $w^j=0$ all derivatives of $f_0$ vanish, and the dimension of the space of solutions to~\eref{Tflat} jumps from 3 to 4. It is this behaviour that we understand as characterising the singularity of the tangent bundle at the conifold.

We can however first deform $\cT_t$ by replacing \eref{Tflat} by the condition
\beq
 V^j\left(\pd{f_t}{w^j} + h_j\right)~=~0\label{Tdef}
\eeq
where the $h_j$ are functions that are not all zero at $w^j=0$ and are not the derivatives of a function~$h$, the latter condition ensuring that we are deforming $\cT_t$ while keeping the underlying space $M_t$ fixed. Then, with the deformed bundle, we may proceed to the limit $f_t\to f_0$ and the space of allowed sections remains 3-dimensional even at the conifold point. The key point is that we have divorced the bundle from being the tangent bundle. We would expect `most' bundles to remain nonsingular at the conifold in the same way as the deformation of the tangent bundle.

We may also think of the equations \eref{Tdef}, as defining sections of a rank three vector bundle on $\Mcheck$. The sections thus defined are independent of position on the $\IP^1$ of the resolution, hence the obtained bundle is trivial as a bundle over the $\IP^1$. This bundle is therefore not a deformation of the tangent bundle of $\Mcheck$. It is clear that the bundle we have defined on $\Mcheck$ deforms back to a bundle on $M_t$. 
\subsection{Deformation of the tangent bundle for the quintic threefold}
We wish to pass from a local to a global description of transgression for compact \cys. Our interest is primarily with the deformations of the tangent bundle of Tian--Yau manifold and its close relatives and the relation of these to bundles on the quotients of the split bicubic. These exhibit certain interesting complications so before examining this case we recall the corresponding situation for the quintic threefold $\IP^4[5]$.  We will first give a standard `explicit' construction~\cite{GSWII} and then a more general account of a type that is more favoured by mathematicians. This has the advantage of being applicable in situations where the explicit construction fails and of generalising in a straightforward way to other bundles. We will repeat this analysis in the following subsection for the Tian--Yau manifold which is a case for which the naive construction fails.

To save writing we denote $\IP^4[5]$ by $Q$ and take the manifold to be defined by a quintic $p(x)$ in the homogeneous coordinates $x_j$, $j=1,\ldots,5$ of the $\IP^4$. The tangent bundle, $\cT_Q$, is the set of pairs $(x,V)$ with $x$ a point of $Q$ and $V=V^j\diff{x^j}$ a vector that now satisfies
\beq
V^j \pd{p}{x^j} = \m\, p~,
\label{quinticV}\eeq
for some $\m$, and is, additionaly, subject to the identification
\beq
V^j + \r\, x^j \simeq V^j    
\label{quinticVid}\eeq
for all $\r$. These conditions are consistent owing to the fact that $p$ is a homogeneous quintic and therefore satisfies the Euler relation $ x^j \pd{p}{x^j} = 5p $. The two conditions reduce the dimension of the space of allowed vectors, at a given point, from 5 to the correct value of 3.

The tangent bundle can be deformed by deforming \eref{quinticV} by quartic polynomials $r_j$ such that
\begin{equation}\label{defquintic}
V^j \left( \pd{p}{x^j} + r_j \right) = \mu\, p \ .
\end{equation}
In order to maintain the identification \eref{quinticVid}, we need to require that
\beq
x^j r_j = 0~ .
\label{defquinticVid}\eeq
We could have imposed the weaker condition $x^j r_j = \m' p$; however an adjustment 
$$
r_j\to r_j-\frac{\m'}{5}\pd{p}{x^j}$$ 
can be made so as to enforce \eref{defquinticVid}. This constraint also prevents the $r_j$ from being the derivatives $\pd{r}{x^j}$ of a quintic $r$; such a deformation would represent a deformation of $Q$, rather than that of $\cT_Q$ on a fixed $Q$.

This construction can be used to estimate the dimension of 
$H^1(Q,\End Q)=H^1(Q,\cT_Q{\otimes}\cT_Q^*)$, the space of first order deformations of 
$\cT_Q$. Having come this far we cannot resist completing the calculation.
Each $r_j$ is a homogeneous polynomial of degree 4 in 5 variables. In these, there are a total of 
$5 \binom{4+5-1}{4} = 350$ parameters. The constraint \eref{defquinticVid}, which, being of degree 5 in 5
variables, imposes $\binom{5 + 5 - 1} {5} = 126$ conditions. We conclude that the number of deformations to $\cT_Q$ is at least $350{ - }126 = 224$.  In this case, in fact $h^1(Q, \cT_Q {\otimes} \cT_Q^*)=224$; for a recent discussion of the deformation of the tangent bundle for the quintic see~\cite{Donagi:2006yf, Li:2004hx}. The quintic is somewhat exceptional: it can happen that the tangent bundle of a manifold has deformations that are not accounted for in this way. 

We now take a more abstract approach, in order to consider deformations of bundles other than the tangent bundle, in particular the direct sum of the tangent bundle and trivial bundles. The adjunction formula~\cite{hart} states that for a projective variety $Y$ embedded in $\IP^n$, the tangent bundle $\cT_Y$ and the normal bundle $\cN_Y$ fit into the short exact sequence
\beq
0 \to \cT_Y \to \left. \cT_{\IP^n}\right|_Y \to \cN_Y \to 0 \ ,
\label{adQ}\eeq
where $\left.\cT_{\IP^n}\right|_Y$ is the tangent bundle of $\IP^n$ 
restricted to $Y$. Next, the Euler sequence on $\IP^n$, restricted to~$Y$,
reads
$$
0 \to \cO_Y \to \cO_{Y}(1)^{\oplus\, n+1} \to \left. \cT_{\IP^n}\right|_Y\to 0 \ ,
$$
and thus, composing the right hand maps in the two sequences, there is a third sequence~\cite{huy}
$$
0 \to \cF \to \cO_{Y}(1)^{\oplus n+1}
\stackrel{f}{\longrightarrow} \cN_Y \to 0 \ ,
$$
where $\cF$ is defined by $\cF = \ker(f)$, in order to make the sequence exact. 
The map $f$ is the analog of~\eref{defquinticVid}. 
For the quintic~$Q$, we get
\beq\label{eulerQ}
0 \to \cF \to \cO_Q(1)^{\oplus 5}
\stackrel{f}{\longrightarrow} \cO_Q(5) \to 0 \ , \qquad
f = \left(\pd{p}{x^j}\right) \ ,
\eeq
with~$f$ now explicitly written as a specific $5 \times 1$ matrix of degree 4 homogeneous polynomials, the partial derivatives of $p$, mapping from $\cO_Q(1)^{\oplus 5}$ to $\cO_Q(5)$. 

Fitting \eref{adQ} and \eref{eulerQ} together yields the following intertwined diagram of short exact sequences:
\beq\label{webQ}
\renewcommand{\arraystretch}{1.1}
\begin{array}{ccccccccc}
&&0&&0&&&& \\
&&\uparrow&&\uparrow&&&& \\
0&\to& \cT_Q &\to&  \cT_{\IP^4}|_Q & \longrightarrow & \cO_Q(5)&\to&0\\
&&\uparrow&&\uparrow&& \vertequals && \\
0&\to& \cF &\to& \cO_Q(1)^{\oplus 5}
&\stackrel{f}{\longrightarrow}& \cO_Q(5) &\to&0\\
&&\uparrow&&\uparrow&&&& \\
&& \cO_Q & = & \cO_Q&&&&\\
&&\uparrow&&\uparrow&&&& \\
&&0&&0&&&& \\
\end{array}
\eeq
The important fact is that the lowest term in the two vertical sequences is the trivial bundle $\cO_Q$ on the quintic.

From \eref{webQ} we see that the bundle~$\cF$ is an extension of $\cT_Q$ by $\cO_Q$, and hence is a deformation of the direct sum $\cT_Q\oplus\cO_Q$ (as explained in~\cite{huy}, $\cF$ is a non-split extension). Deformations of $\cF$ are obtained by deforming the map~$f$, given by the partial derivatices $\left(\pd{p}{x^j}\right)$, to a map $ \left(\pd{p}{x^j}+ r_j\right)$, for some quatrics $r_j$ as in~\eref{defquintic}. Imposing~\eref{defquinticVid} corresponds, in this language, to the statement that the deformed~$f$ maps the distinguished subbundle $\cO_Q$ of $\cO_Q(1)^{\oplus 5}$ to zero, and thus
we obtain a deformation of $\cT_Q$. If~\eref{defquinticVid} fails, 
then we obtain a deformation of ${\cal F}$, and hence of $\cT_Q\oplus\cO_Q$.
\subsection{Deforming the tangent bundle of the Tian--Yau manifold}
Let us move on to the manifold of our principal interest here.  It is instructive to first run through the explicit construction of the deformation of the tangent bundle to see how this fails.

To avoid possible complications related to the
resolution of the fixed point sets of $B{\times}C$ let us focus on $M$, presented as
$$
\def\ns{\hskip-3pt}
\kbordermatrix{&\ns F\ns&\ns G\ns&\ns H\ns\\
      x~~~\IP^3&1&3&0\\
      y~~~\IP^3&1&0&3}\quotient{A}~~=~~M
$$ 

There are now 3 defining equations, which we will also denote by $p^\a=(F,G,H)$. We can think of the tangent vector as a differential operator
$$
V = X+Y \ ;
\qquad
X=X^a\frac{\partial}{\partial x^a} \ , \quad
Y=Y^b\frac{\partial}{\partial y^b} \ ,
$$
where the indices $a,\, b=0,\ldots,3$ run through the two $\IP^3$'s. We will also, as previously, have need of indices $j,\, k$ that run over the range $1,2,3$, and which are understood to take values modulo 3.

The analogue to \eref{defquintic}, requiring $V$ to be tangent to $M$, is the condition
\beq\label{vec}
V(p^\a)=m^\a{}_\b\, p^\b \ , 
\eeq
where now $m^\a{}_\b$ is in general a matrix of polynomials but which in this case, owing to the degrees of the polynomials, is a diagonal matrix of constants. The Euler relations now read
$$
x^{a}\pd{p^\a}{x^a}=\text{deg}_{x}(p^\a)\, p^\a \ , \qquad
y^{b}\pd{p^\a}{y^b}=\text{deg}_{y}(p^\a)\, p^\a \ ,
$$
where here $\text{deg}_{x}(p^\a)$ denotes the degree of $p^\a$ as a function of $x$ and similarly for
$\text{deg}_{y}(p^\a)$ . We also identify
\beq
(X^{a}+\rho x^{a},\, Y^{b}+\sigma y^{b})\simeq (X^{a},\, Y^{b}) \ .
\label{idM}\eeq

the tangent bundle $\cT_M$ is now deformed by requiring that $V=X+Y$ satisfy
\beq
X^{a}\left(\pd{p^\a}{x^a} + r^\a{}_{a}\right) + Y^{b}\left(\pd{p^\a}{y^b} + s^\a{}_{b}\right)
~=~m^\a{}_\b\, p^\b \ ,
\label{XYdef}\eeq
where the quantities $r^{\alpha}{}_{a}$ and $s^{\alpha}_{b}$ are
matrices of polynomials of the same degrees as $\pd{p^\a}{x^a}$ and $\pd{p^\a}{y^{b}}$, respectively
while
$m^{\alpha}{}_{\beta}$ is a matrix of constants. In order to maintain the identification \eref{idM}, we write 
$$
R^\a{}_a~=~\pd{p^\a}{x^a} + r^\a{}_{a}~~~\text{and}~~~
S^\a{}_b~=~\pd{p^\a}{y^b} + s^\a{}_{b}
$$
and require
\beq\label{defM}
x^{a}R^{\alpha}{}_{a}=n^{\alpha}{}_{\beta}\, p^{\beta}~, \qquad
y^{b}S^{\alpha}{}_{b}={\tilde n}^{\alpha}{}_{\beta}\, p^{\beta}
\eeq
for some constant matrices $n^{\alpha}{}_{\beta}$, ${\tilde n}^{\alpha}{}_{\beta}$.

Note first that, since the second defining equation, $p^2=G$, has zero degree in $y^a$ and similarly, $p^3=H$ has zero degree in $x^a$, the quantities $S^{2}{}_{b}$ and $R^{3}{}_{a}$  both vanish.
A little thought then shows that the most general form for $r^2{}_a$ that is invariant under $A$ and satisfies 
\eref{defM} has the form
$$
r^{2}{}_{a}dx^{a} = \sum_{j}\left(\k\, x_{0} + \l\, x_{j} +
\sum_{\pm}\m_{\pm}\, x_{j\pm 1}\right)\left(x_{j}dx_{0} - x_{0}dx_{j}\right) +
\sum_{j}\sum_{\pm}\n_{\pm}(x_{j\pm 1}^{2} -x_{j}x_{j\mp 1})dx_{j} \ .
$$
and contains 6 parameters. A similar expression holds for $s^3{}_b$.

A problem, however, arises in relation to the deformation corresponding to $p^1=F$. This polynomial is bilinear so the conditions \eref{defM}, applied to this case, force the relations
$$
R^1{}_a~=~u\,\pd{F}{x^a}~~~\text{and}~~~S^1{}_b~=~v\,\pd{F}{y^b}
$$
for some constants $u$ and $v$. Our previous observation concerning transgression of deformation of tangent bundles, as in \eref{Tdef}, was that once the tangent bundle has been deformed sufficiently generally, then the analogs of our quantities $R^\a{}_a$ and $S^\a{}_b$ are all nonzero at the conifold and prevent the bundle from becoming singular. Here this fails since at the conifold (as in~\eref{conifoldF}) $F$ is necessarily of the form $F_0=UV-WZ$, for some polynomials $U,V,W,Z$, that vanish at the nodes. Thus all the derivatives of $F$ vanish at the nodes and hence also the quantities $R^1{}_a$ and $S^1{}_b$. So in this case, the deformation does not help to avoid the singularity associated with the tangent bundle at the conifold point.

We can however save the situation by abandoning the identification in \eref{idM}. Our bundle is then no longer a deformation of the tangent bundle of $M$, but a deformation of $\cT_M\oplus\cO\oplus\cO$. Alternatively we could abandon just one of the identifications of \eref{idM}, in which case the bundle becomes a deformation~of~$\cT_M\oplus\cO$.

We can also study deformations in the language of intertwined short exact sequences. As we have seen, the key equation is~$F$; we take $M^F$ as the (5-dimensional) hypersurface given by the vanishing of $F$ in the ambient space $\IP^3{\times}\IP^3$ (we can impose the other equations later). In complete analogy with~\eref{webQ}, we have
\beq\label{webF}
\renewcommand{\arraystretch}{1.2}
\begin{array}{ccccccccc}
&&0&&0&&&& \\
&&\uparrow&&\uparrow&&&& \\
0&\to& \cT_{M^F} &\to&  \cT_{\IP^3{\times}\IP^3}|_{M^F}
 &\longrightarrow& \cO_{M^F}(1,1)&\to&0\\
&&\uparrow&&\hphantom{\vph}\!\uparrow\!\vph && \vertequals && \\
0&\to& \cF &\to& \cO_{M^F}(1,0)^{\oplus 4} \oplus \cO_{M^F}(0,1)^{\oplus 4}
&\stackrel{f}{\longrightarrow}& \cO_{M^F}(1,1) &\to&0\\
&&\uparrow&&\uparrow&&&& \\
&& \cO_{M^F}^{\oplus 2}  & = & \cO_{M^F}^{\oplus 2} &&&&\\
&&\uparrow&&\uparrow&&&& \\
&&0&&0&&&& \\
\end{array}
\eeq
The important facts to notice are the following. The map $f$ is now given by an $8 \times 1$ matrix with 4 entries of bidegree $(1,0)$ and 4 entries of bidegree $(0,1)$, corresponding to the operators $\diff{x^a}$ and $\diff{y^a}$. The kernel of the map $\varphi$ now is actually $\cO_{M^F}^{\oplus 2}$, two copies of the trivial bundle on ${M^F}$. Therefore the bundle $\cF$, which is the kernel of $f$, and hence whose deformations are measured by \eref{XYdef}, is a deformation of $\cT_{M^F}\oplus \cO_{M^F}^{\oplus 2}$, a rank 5 bundle. Varying the map $f$ results in further deformations of this bundle. If either $u$ or $v$ vanished, then we would be deforming the rank 4 bundle $\cT_{M^F} \oplus \cO_{M^F}$; in case the both vanish, we are back to deformations of $\cT_{M^F}$.

It is interesting that, on the split bicubic side, the motivation for studying bundles of higher rank derives from the fact that the commutant of the background gauge group in $E_8$ is the gauge group of the low energy effective theory. For the `standard embedding', the bundle associated with the background gauge group is identified with the tangent bundle, the background gauge group is  $SU(3)$ and the gauge group of the low energy effective theory is $E_6$, the commutant of $SU(3)$ in $E_8$. A bundle of higher rank can have a larger background gauge group leading to a gauge group for the low energy theory that is a subgroup of $E_6$. In particular a rank 4 bundle on the \cym\ can give rise to a gauge group $SO(10)$ in spacetime and a rank 5 bundle can give rise to a gauge group $SU(5)$. Bundles of higher rank are therefore of interest because they can lead to appealing phenomenology. It is curious that we are here driven to rank 5 bundles by a wish to transgress.
\subsection{On the geometry of $X^{3,3}$ and the heterotic vector bundle it carries}
The manifold $X^{3,3}$ is a $\IZ_3{\times}\IZ_3$ quotient of the split bicubic $X^{19,19}$. 
The latter is the fiber product of two ninth del Pezzo ($dP_9$) surfaces~\cite{Braun:2005nv}, 
$$
\cicy{t~~\IP^1\\ \x~~\IP^2}{1\\3\\}~=~B_1~~~~~\text{and}~~~~~
\cicy{t~~\IP^1\\ \eta~~\IP^2}{1\\3\\}~=~B_2~.
$$
The equations for the split bicubic $X^{19,19}$ are of the form
\bean
F^1&=& t_1\, U(\x) + t_2\, W(\x)\\[1ex]
F^2&=& t_1\, Z(\eta) + t_2\, V(\eta)
\eean
with $U$ and $W$ cubics in $\x$, and $V$ and $Z$ cubics in $\eta$. Particular equations invariant under $A{\times}D$ are given by~\eref{ADcubics}. The surfaces $B_1$ and $B_2$ correspond to the vanishing of $F^1$ and $F^2$ respectively. Consider the surface $B_1$. The two cubics $U(\x)$ and $V(\x)$ vanish simultaneously in nine points and for each $t$ the locus $F^1=0$ is a cubic curve, $E_1(t)$, in $\IP^2$ which passes through these nine points. It is easy to see that for each $\x\in\IP^2$, that is not one of these nine special points, there is a unique $t\in\IP^1$ such that $E_1(t)$ contains $\x$. As $t$ varies $E_1(t)$ sweeps out the surface $B_1$. Apart from the nine special points, each point of $B_1$ derives from a unique point of $\IP^2$. Consider the map that projects the points of $B_1$ back onto the corresponding points of $\IP^2$ in this way. For $\x_*$ one of the special points, the equation $F^1=0$ is satisfied for every $t$. Thus an entire $\IP^1$ of $B_1$ projects down to $\x_*$. We shall refer to the $\IP^1$'s that project down to the nine special points as exceptional lines. We have just worked through the classical fact that the surfaces $B_i$ above are $dP_9$'s, surfaces obtained by blowing up $\IP^2$ in nine points. Along with the birational morphims $B_i\to \IP^2$ just discussed, there are also the projections $B_i\to \IP^1$, which exhibit the $dP_9$'s as elliptic fibrations. 

We turn now to a description of the homology~\cite{Braun:2004xv} of the quotient manifold $X^{3,3}$. The space $H^2(X^{3,3},\IC)=H^2(X^{19,19}, \IC)^{\ZZ}$ is three-dimensional, with basis 
$(\tau_1, \tau_2, \phi)$. Here the $\tau_{i}$ are the pull-backs of the $\zz$-invariant divisors in $B_{i}$, the hyperplane classes pulled back from~$\IP^2$. The class $\phi$ is the pull-back of the common fiber class $\mathfrak f$, the fibre of the projection of the threefold to $\IP^1$. The intersection numbers are
\beq
\begin{array}{r@{{}={}}lr@{{}={}}l}
\tau_i^3 &0 &\phi\, \tau_i & 3\, \tau_i^2~;\qquad i=1,2 \ ;\\[1ex]
\tau_1^2 \tau_2 & \tau_1 \tau_2^2 = 3~, &~~~~\ph^2&0~;\\
\end{array}
\label{inter}\eeq
in particular, $H^4(X^{3,3},\IC)=H^4(X^{19,19}, \IC)^{\ZZ}$ is spanned by the classes
$(\tau_1^2, \tau_2^2, \tau_1\tau_2)$.

The stable $\zz$-equivariant $SU(4)$ bundle $\cV$ on $X^{3,3}$ which gives the MSSM 
spectrum~\cite{Braun:2005nv} descends from an $\zz$-equivariant bundle $\Vt$ on $X^{19,19}$. The bundle $\Vt$ in turn is given by the short exact sequence
\beq
0\longrightarrow \cV_1 \longrightarrow \Vt \longrightarrow \cV_2 \longrightarrow 0
\label{Vtdef}\eeq
of two rank $2$ bundles, $\cV_1$ and $\cV_2$. Each of these is itself the tensor
product of a line bundle with a rank $2$ bundle pulled back from a
$dP_9$ factor of $X^{19,19}$:
$$
\cV_1 = \cO_X(-\tau_1+\tau_2) \otimes \pi_1^\ast(\cW_1)
\,, \quad
\cV_2 = \cO_X(\tau_1-\tau_2) \otimes \pi_2^\ast(\cW_2) \ ,
$$
where
\bean
&&
0 \longrightarrow \a_1\, \cO_{B_1}(- {\mathfrak f}) \longrightarrow \cW_1 \longrightarrow
\a_1^2\, \cO_{B_1}( {\mathfrak f}) \otimes I_3^{B_1} \longrightarrow 0 \ , \\[2ex]
&&0 \longrightarrow \a_2\, \cO_{B_2}(-{\mathfrak f})  \longrightarrow
\cW_2 \longrightarrow \a_2^2\, \cO_{B_2}({\mathfrak f}) \otimes I_6^{B_2} \longrightarrow 0~;
\eean
here $I_3^{B_1}$ and $I_6^{B_2}$ denote the ideal sheaf of 3 and 6 points in $B_1$ and $B_2$ respectively, and $\a_1$ and $\a_2$ are third roots of unity. The Chern classes of $\Vt$ can be computed from these sequences to be 
\beq
c_1(\Vt) = 0 \ , \quad
c_2(\Vt) = \tau_1^2 + 4 \tau_2^2 + 4 \tau_1 \tau_2 \ , \quad
c_3(\Vt) = -54~.
\label{chernVPenn}
\eeq

\subsection{Transgression between $N^{\,2,11}$ and $X^{3,3}$}
Recall from \SS2.4 (compare Table~2) that there is a conifold transition between the quotient $X^{3,3}$ of the split bicubic, and the manifold $N^{\,2,11}$. We wish to investigate the possibility that the heterotic bundle discussed above can arise as a transgression under this transition.
Before we do that, we make one further remark concerning the relation between exceptional lines on the $dP_9$'s to the conifold lines of the split bicubic $X^{19,19}$, the $\IP^1$'s of the split bicubic that project down to the nodes of the conifold. We have seen that the conifold lines correspond to the solutions of 
$$
U(\x)=W(\x)=0~~~~\text{and}~~~~V(\eta)=Z(\eta)=0~.
$$
These equations are solved precisely where $(\x,\,\eta)=(\x_*,\,\eta_*)$. In this way we see that the 81 conifold lines of the split bicubic are in correspondence to the 81 pairs of exceptional lines on $B_1{\times}B_2$. 

\subsubsection{Triviality of the heterotic bundle on conifold lines}
A necessary condition for a bundle to be a transgression through a conifold is that the bundle be trivial on the conifold lines. For the present case we denote the homology class of the conifold lines of the quotient $X^{3,3}$ by $L\in H^4(X^{3,3},\IC)$. 

The nonsingular manifold $N^{\,2,11}$ has a two-dimensional cohomology group $H^2(N^{\,2,11}, \IC)$ spanned by classes $H_1$ and $H_2$, the two hyperplane classes restricted from $\IP^2{\times }\IP^2$. The hyperplane classes also give cohomology classes in $H^2(N^{\,2,11}_0)$, the conifold degeneration. On the other hand, we have an inclusion $H^2(N^{\,2,11}_0, \IC)\hookrightarrow H^2(X^{3,3}, \IC)$, where $X^{3,3}$ is the split manifold after the conifold transition; under this map, the classes $H_i$ map to the classes $\tau_i$ for $i = 1,2$, the latter being hyperplane sections from the two individual $\IP^2$'s.
Indeed, for the manifold $N^{\,2,11}$ we have the intersection numbers
$$
H_1^3 = H_2^3 = 0~,~~~
H_1^2 H_2 = H_2^2 H_1 = 3 \ .
$$
which agrees with \eref{inter} under the identification $H_i\mapsto \tau_i$.

Since the classes $\tau_i$ arise as pullbacks from $N^{\,2,11}_0$ under the resolution map 
$X^{3,3}\to N^{\,2,11}_0$, we know by the projection formula that the curve class $L\in H^4(X^{3,3},\IC)$ must be perpendicular to $\tau_1$ and $\tau_2$. Hence we can write $L$ in terms of our homology basis as 
\beq
L = - \tau_1^2 -  \tau_2^2 +  \tau_1 \tau_2 \ , 
\label{Lexpr}\eeq
where we also used the condition that since $\phi$ is an effective class on $X^{3,3}$, 
we must have $\phi\cdot L\geq 0$. 

We learn about the restriction of $\Vt$ to a conifold curve $L\cong\IP^1\subset X^{19,19}$ 
in the cover by restricting the exact sequence \eref{Vtdef} to get
$$
0 \to  \Big(\cO_X(-\tau_1+\tau_2) \otimes \pi_1^\ast(\cW_1)\Big)\lower3pt\hbox{$|_L $}\to
\Vt|_L \to \Big(\cO_X(\tau_1-\tau_2) \otimes \pi_2^\ast(\cW_2)\Big)\lower3pt\hbox{$|_L $} \to 0~.
$$
Now, from \eref{inter} and \eref{Lexpr}, we see that 
$$
\cO_X(-\tau_1+\tau_2)|_L ~\cong~ \cO_L(\-\tau_1^2 \tau_2 + \tau_1 \tau_2^2) ~\cong~ \cO_L
$$
and similarly for $\cO_X(-\tau_1+\tau_2)|_L$. Hence,
\beq\label{VtL}
0 \to  \pi_1^\ast(\cW_1)|_L \to \Vt|_L \to  \pi_2^\ast(\cW_2)|_L \to 0 \ .
\eeq

On the other hand, we claim that both restrictions $\pi_i^\ast(\cW_i)|_L $ are trivial. From the definition of $\cW_i$,
$$
0 \to \cO_{B_i}(- {\mathfrak f})|_L \to \cW_i|_L \to \cO_{B_i}({\mathfrak f})|_L \to 0 \ ,
$$
where we used the fact that the points which define the ideal sheaves $I_3$ and $I_6$ can be moved away from  $L$. Thus the extension class of $\cW_i|_L$ lives in 
$$
\text{Ext}^1\Big(\cO_{B_i}({\mathfrak f})|_L, \cO_{B_i}(-{\mathfrak f})|_L \to 0\Big) 
\cong H^1\big(L, \cO_{L}(2 {\mathfrak f}|_L)\big) = H^1\big(\IP^1, \cO_{\IP^1}(6)\big) \ne 0$$ 
since ${\mathfrak f}\cdot\p_i( \t_i )= 3$. The generic element of this Ext-group corresponds to the extension $\cW_i|_L \cong\cO_{\IP^1}^2$; since the extensions defining the $\cW_i$
can be chosen to be generic, we are dealing here with a generic extension as well. 
Substituting into \eref{VtL}, we conclude that 
$$
\Vt|_L \cong \cO_X^{\oplus 4}
$$
on the conifold curves $L\subset X$.
\subsubsection{Candidate bundles}

We now approach the transition from the other side, attempting to construct some 
bundles on the manifold $N^{\, 2,11}$ with the right Chern classes. 
The first salient feature of the bundle $\Vt$ is that $c_3(\Vt)=-6$ on $X^{3,3}$. The bundle descends from an equivariant bundle on the covering space~$X^{19,19}$ with $c_3 = -54$. For the transgression we should seek a bundle with $c_3=-6$ on $N^{\,2,11}$ or an equivariant bundle also with $c_3 = -54$ on the~bicubic $N^{\,2,83}$.

There has been some recent revival in interest in CICY's, especially in constructing bundles over 
them~\cite{monad}. A large class of bundles can be constructed over algebraic manifolds via the monad construction. Over projective spaces, for example, all bundles arise this way. Let us try, therefore, to construct a monad-type bundle, as the kernel of bundle maps between sums of line-bundles, over the bicubic. Such bundles   $\cV$ are defined by the short-exact sequence
$$
0 \to \cV \to \cB \stackrel{f}{\longrightarrow} \cC \to 0 \ ,
$$
with
\beq
\cB = \bigoplus_{i=1}^{r_\cB} \cO_X (b_1^i, b_2^i) \ , \quad
\cC = \bigoplus_{j=1}^{r_\cC} \cO_X(c_1^j, c_2^j) \ .
\label{monad}
\eeq
where $\cO(a,b)$ is the line bundle of bi-degree $(a,b)$ on the bicubic. In terms of the hyperplane classes, this line bundle is to be thought of as $\cO(a H_1, b H_2)$. The $b^i_s$ and $c^j_s$ are integers such that for $s=1,2$ and all $i$, $j$, $c_s^j \ge b_s^i \ge 0$ in order that $f$ be a well-defined polynomial map. In particular, $\cV$ has the following topological properties
($r,s,t=1,2$):
\bea
\nn \mbox{rk}(\cV) &=& r_\cB - r_\cC \ , \\
\nn [c_1(\cV)]_s &=& \sum_{i=1}^{r_\cB} b_s^i - \sum_{j=1}^{r_\cC} c_s^j \ ,\\
\nn [c_2(\cV)]_r &=& \frac12 \sum_{s,t=1}^{k} y^{\mbox{\scriptsize top}}_{rst} 
   \left(\sum_{j=1}^{r_\cC} c_s^j c_t^j- \sum_{i=1}^{r_\cB} b_s^i b_t^i \right) \ , \\ \label{chernV} 
c_3(\cV) &=& \frac13 \sum_{r,s,t=1}^{k} y^{\mbox{\scriptsize top}}_{rst} 
   \left(\sum_{i=1}^{r_\cB} b_r^i b_s^i b_t^i - \sum_{j=1}^{r_\cC} c_r^j c_s^j c_t^j \right) \ ,
\eea
where $y^{\mbox{\scriptsize top}}_{rst}$ is the intersection form on the bi-cubic. Finally, we must descend this bundle onto the $A {\times} D$ quotient~$X^{3,3}$. This can be done by finding an $A {\times} D$-equivariant map.

It is now a matter of solving for non-negative integer values for $b_s^i$ and $c_s^j$ such that first of all
$c_1(\cV) = 0$ and $c_3(\cV) = - 54$. Then, in the $(H_1, H_2)$ basis, we can form the integer 2-vector 
$c_2(\cV) \cdot (H_1, H_2)$. This should then be compared, with $c_2(\Vt)$ expressed in the $(\tau_1, \tau_2)$ basis. Examining \eref{chernVPenn}, we see that in this basis
$$
c_2(\Vt)~=~(\tau_1^2 + 4 \tau_2^2 + 4 \tau_1 \tau_2).(\tau_1,\, \tau_2) = (24,\,15)~.
$$

After performing a computer scan for possible solutions, we find
the following rank 4 candidate monad bundles with all the above requisite
properties:

$$
\def\str{\vrule height16pt width0pt depth8pt}
\begin{tabular}{| c | c |}
\hline \str $\cB$ & $\cC$\\
\noalign{\hrule\vskip3pt}
\hline \str $\cO(0,1)^{\oplus 3} \oplus \cO(0,2) \oplus \cO(1,0)^{\oplus 2}$ & $\cO(1,2) \oplus \cO(1,3)$\\
\hline \str $\cO(0, 1)^{\oplus 3} \oplus \cO(1, 0) \oplus \cO(1, 1)^{\oplus 3}$ 
               & $~\cO(1, 2) \oplus \cO(1, 3) \oplus \cO(2,1)~$\\
\hline \str $\cO(0,1)^{\oplus 3} \oplus \cO(1,1)^{\oplus 6}$ & $\cO(1, 2)^{\oplus 4} \oplus \cO(2,1)$\\
\hline \str $~~\cO(0, 1)^{\oplus 3} \oplus \cO(0,2)  \oplus \cO(1,0) \oplus \cO(1, 1)^{\oplus 3}~$ 
               & $\cO(1,2)^{\oplus 4}$\\
\hline
\end{tabular}
$$
\vskip5pt
In this section we have, inspired by the intimate relation between the Tian-Yau and the bicubic sequences of manifolds, discussed in detail how one may transgress a bundle from a manifold to another through a conifold transition. Though the transgressions of bundles are old ideas in principle, here we have considered an explicit example. Our construction predicts a bundle on $N^{\,2,11}$, which is a transgression of the heterotic
MSSM bundle on $X^{3,3}$ and which may afford here a simpler description. This is an intriguing investigation to which we intend to return in a future publication.
\newpage
\section{Concluding Speculations}
We began by plotting discretely different \cys\ and were naturally drawn to the fact that each manifold has a parameter space and these spaces meet in loci corresponding to certain singular manifolds. For the cases we have been considering the relationship between the manifolds is close and the singular manifolds in which the parameter spaces meet are the conifolds which are only mildly singular. It is an old speculation~\cite{Reid} that the space of all \cys\ may be connected in this way. 

At a technical level it is known that the parameter spaces of a great many \cys\ form a connected web. For example it is known that all CICY's are connected by a series of conifold 
transitions~\cite{Green:1988bp, hubsch}.
It is known also that the parameter spaces of all the manifolds of the Kreuzer-Skarke list form a 
connected web~\cite{Avram:1997rs}, which is moreover connected to the web of CICY's, though the singular manifolds can be more singular than conifolds. Much work has been done on string theory compactified on conifolds and related spaces subsequent to~\cite{Strominger:1995cz} and in many cases conifold transitions seem to be physically acceptable. Nevertheless it is not known to what extent it is possible for the universe to move from one \cym\ to another. However if it is possible in heterotic string theory then transgression will be an important part of the story.

Looking at the tip of the landscape it is hard not to speculate that there may be a physical mechanism allowing transitions between what appear classically to be different vacua thereby permitting the universe to trickle down to a very special corner of the landscape, an oasis where only very few \cys\ reside.
\vskip1in
\leftline{\large\bf Acknowledgements}
It is a pleasure to thank Maximilian Kreuzer for extensive discussions and for sharing his results prior to publication and to acknowledge fruitful conversations with Lara Anderson, Andre Lukas, Graham Ross, David Skinner and Duco van Straten. We are grateful also to Rosalind Thomas for erudite instruction in matters of Greek etymology and to Raquel Candelas for instruction in Latin. PC also wishes to thank Pedro Ferreira for encouragement and discussions of this work while running many miles. YHH is indebted to the gracious patronage, through the FitzJames Fellowship, of Merton College, Oxford. The research of BS is partially supported by OTKA grant K61116.
\newpage

\end{document}